\documentclass[conference]{IEEEtran}
\usepackage{cite}
\usepackage{amsmath,amssymb,amsfonts}
\usepackage{graphicx}
\usepackage{textcomp}
\usepackage{xcolor}
\usepackage[hyphens]{url}
\usepackage{algorithm}
\usepackage{algorithmicx}
\usepackage{algpseudocode}

\def\BibTeX{{\rm B\kern-.05em{\sc i\kern-.025em b}\kern-.08em
    T\kern-.1667em\lower.7ex\hbox{E}\kern-.125emX}}

\title{
\thiswork{}: A Fast and Flexible \\ Collective Communication Framework for \\ Commodity Processing-in-DIMM Devices} 

\usepackage{booktabs}

\usepackage{xcolor}

\usepackage{amsmath,amsfonts}
\usepackage{amssymb}

\definecolor{olivegreen}{rgb}{0, 0.6, 0}
\definecolor{redorange}{HTML}{FF5349}
\definecolor{blue(ncs)}{rgb}{0.0, 0.53, 0.74}
\definecolor{navy}{HTML}{273BE2}
\definecolor{cerulean}{HTML}{2a52be}

\definecolor{black}{HTML}{000000}
\definecolor{white}{HTML}{ffffff}
\definecolor{color1}{HTML}{ACE5EE}
\definecolor{color2}{HTML}{0093AF}
\definecolor{color3}{HTML}{CC0000}
\definecolor{color4}{HTML}{0087BD}
\definecolor{color5}{HTML}{333399}
\definecolor{color6}{HTML}{20B2AA}

\usepackage{xspace}

\usepackage{graphicx}
\usepackage[export]{adjustbox}
\usepackage{wrapfig}

\usepackage{verbatim}

\usepackage{multicol}
\usepackage{multirow}
\usepackage{makecell}

\usepackage{nicefrac}

\usepackage{algorithm}
\usepackage{algpseudocode}
\algnewcommand{\IfThen}[2]{
  \State \algorithmicif\ #1\ \algorithmicthen\ #2\ }
\algnewcommand{\IfThenElse}[3]{
  \State \algorithmicif\ #1\ \algorithmicthen\ #2\ \\ \algorithmicelse\ #3\ }

\usepackage{bookmark}

\usepackage{soul}

\usepackage[noabbrev, capitalize]{cleveref} 
\Crefname{section}{\S}{\S}

\usepackage{tcolorbox}
\usepackage{pifont}
\newcommand{\cmark}{{\color{olivegreen}\ding{51}}}

\usepackage{tikz}
\newcommand*\circled[1]{\tikz[baseline=(char.base)]{
            \node[shape=circle,draw,inner sep=0.4pt] (char) {#1};}}

\usepackage{enumitem}
\setlist{leftmargin=3.5mm}

\newcommand{\asplossubmissionnumber}{319}
\newcommand{\thiswork}{PID-Comm\xspace}

\newcommand{\pr}{PE-assisted reordering\xspace}
\renewcommand{\Pr}{PE-assisted reordering\xspace}
\newcommand{\im}{in-register modulation\xspace}
\renewcommand{\Im}{In-register modulation\xspace}
\newcommand{\cm}{cross-domain modulation\xspace}
\newcommand{\Cm}{Cross-domain modulation\xspace}

\newcommand{\x}{\texttimes\xspace}

\newcommand{\JL}[1]{{\color{magenta}[\textbf{\sc JLee}: \textit{#1}]}}

 \iffalse 
\renewcommand{\JL}[1]{}
\else
\fi

\newcommand{\JU}[1]{{\color{blue}[\textbf{\sc JU}: \textit{#1}]}}
\newcommand{\SY}[1]{{\color{cerulean}[\textbf{\sc SY}: \textit{#1}]}}

\newcommand{\revref}[1]{\hyperref[rev:#1]{\color{magenta}#1}}

\newcommand{\figref}[1]{\hyperref[fig:#1]{\color{blue}\figurename~\ref{fig:#1}}}
\newcommand{\tblref}[1]{\hyperref[tbl:#1]{\color{blue}\tablename~\ref{tab:#1}}}
\newcommand{\secref}[1]{\hyperref[sec:#1]{\color{blue}Section~\ref{sec:#1}}}
\newcommand{\algoref}[1]{\hyperref[alg:#1]{\color{blue}Algorithm~\ref{alg:#1}}}

\hypersetup{hidelinks}

\usepackage[normalem]{ulem}
\usepackage{soul}

\newcommand{\scheme}[0]{\thiswork{}} 

\begin{document}

\pagenumbering{empty}

\include{pythonlisting}


\author{%
\IEEEauthorblockN{%
Si Ung Noh\IEEEauthorrefmark{4}\textsuperscript{,1},
Junguk Hong\IEEEauthorrefmark{4}\textsuperscript{,1},
Chaemin Lim\IEEEauthorrefmark{2},
Seongyeon Park\IEEEauthorrefmark{4},
Jeehyun Kim\IEEEauthorrefmark{2},\\
Hanjun Kim\IEEEauthorrefmark{3},
Youngsok Kim\IEEEauthorrefmark{2}\textsuperscript{,\P{}}
and Jinho Lee\IEEEauthorrefmark{4}\textsuperscript{,\P{}}}\vspace{2pt}%
\IEEEauthorblockA{%
\IEEEauthorrefmark{4}\textit{Department of Electrical and Computer Engineering, Seoul National University}\\%
\IEEEauthorrefmark{2}\textit{Department of Computer Science, Yonsei University}\\%
\IEEEauthorrefmark{3}\textit{School of Electrical and Electronic Engineering, Yonsei University}}
\vspace{3pt}%
\IEEEauthorblockA{%
\{siung98, junguk16\}@snu.ac.kr, cmlim@yonsei.ac.kr, syeonp@snu.ac.kr, jeehyun990@yonsei.ac.kr,\\
\{hanjun, youngsok\}@yonsei.ac.kr, leejinho@snu.ac.kr}}

\maketitle
\thispagestyle{plain}
\pagestyle{plain}
\pagenumbering{arabic}
\setcounter{section}{0}


\begin{abstract}
Recent dual in-line memory modules (DIMMs) are starting to support processing-in-memory (PIM) by associating their memory banks with processing elements (PEs), allowing applications to overcome the data movement bottleneck by offloading memory-intensive operations to the PEs.
Many highly parallel applications have been shown to benefit from these PIM-enabled DIMMs, 
but further speedup is often limited by the huge overhead of inter-PE collective communication.
This mainly comes from the slow CPU-mediated inter-PE communication methods which incurs significant performance overheads, making it difficult for PIM-enabled DIMMs to accelerate a wider range of applications. 
Prior studies have tried to alleviate the communication bottleneck, but they lack enough flexibility and performance to be used for a wide range of applications.

In this paper, we present \textbf{\scheme{}}, a fast and flexible inter-PE collective communication framework for commodity PIM-enabled DIMMs.
The key idea of \scheme{} is to abstract the PEs as a multi-dimensional hypercube and allow multiple instances of inter-PE collective communication between the PEs belonging to certain dimensions of the hypercube.
Leveraging this abstraction, \scheme{} first defines eight inter-PE collective communication patterns that allow applications to easily express their complex communication patterns.
Then, \scheme{} provides high-performance implementations of the inter-PE collective communication patterns
optimized for the DIMMs.
Our evaluation using 16 UPMEM DIMMs
and 
representative parallel algorithms
shows that \scheme{} greatly improves the performance by up to 5.19\texttimes{} compared to the existing inter-PE communication implementations.
The implementation of PID-Comm is available at \url{https://github.com/AIS-SNU/PID-Comm}.
\end{abstract}

\footnotetext[1]{%
Co-first authors\\%
\indent\textsuperscript{\P{}}Co-corresponding authors}

\section{Introduction}
\label{sec:intro}

%
Processing-in-memory (PIM) has emerged as a promising solution to overcome the data movement bottleneck caused by frequent memory accesses of memory-intensive applications~\cite{diva, tesseract, googleworkloads}.
With the introduction of commodity dual in-line memory modules (DIMMs) supporting PIM (e.g., UPMEM DIMMs~\cite{upmem}), recent studies have shown that various memory-intensive applications can be greatly accelerated on real systems~\cite{gapim, prim, sparsep, pidjoin, transpimlib}. 
PIM-enabled DIMMs represent one form of PIM, realized by associating in-DIMM processing elements (PE) to each of their memory banks.

%
Unfortunately, the effectiveness of PIM-enabled DIMMs still remains highly limited despite achieving large performance improvements with highly parallel applications.
One important limitation of the commodity PIM-enabled DIMMs is the huge overhead of inter-PE communication.
From the lack of direct inter-PE paths in the modern DRAM architecture, the host CPU becomes the medium for transferring data between PEs~\cite{prim, mcn}, incurring large performance overheads upon inter-PE communication.
Some approaches attempt to utilize off-chip channels for different purposes~\cite{abcdimm}, or even dedicated links~\cite{dimmlink, bioaim}.
However, such support is difficult to be realized in the near future due to strict timing, chip area, and power constraints imposed on DIMMs~\cite{pimulator}.

The current methods for inter-PE communication are not only slow but also lack an adequate abstraction model. 
This not only poses a great burden to the programmers but also leads to low achievable throughput for communications.
These issues greatly limit the benefits of PIM-enabled DIMMs and have made prior studies mainly focus on accelerating highly parallel applications with little to no inter-PE communications. 

%

Some approaches addressed these inter-PE communication issues~\cite{upmemtraining, pidjoin, simplepim}, but they are either limited in flexibility and performance~\cite{simplepim} or confined to application-specific workarounds~\cite{upmemtraining, pidjoin}.
Therefore, to fully exploit the high potential of PIM-enabled DIMMs, there is a clear need for a fast and flexible inter-PE communication framework that supports diverse and complex communication patterns and implementations tailored for the PIM-enabled DIMMs.

%
In such circumstances, we present \emph{\scheme{}}, a collective communication framework as a promising solution for fast and flexible inter-PE collective communication on commodity PIM-enabled DIMMs.
\scheme{} abstracts the PEs as a user-defined multi-dimensional virtual hypercube, and maps them to the hierarchy of the DRAM architecture.
Using the hypercube, users can invoke multi-instance collective communication~\cite{mpitharkur} with high performance over various dimensions.

To realize the virtual hypercube model and provide high performance, 
\thiswork uses a set of novel techniques to mitigate bottlenecks in the conventional approaches.
Specifically, we devise algorithmic tricks such that 1) the computational burden of the host is reduced, 2) host memory access is eliminated, and 3) domain transfer between PIM and host is avoided, while providing flexible abstractions to the user.


We implement \scheme{} on a real system equipped with 16 UPMEM DIMMs~\cite{upmem}.
Our evaluation reveals that \scheme{} achieves 
speedup up to 5.19\texttimes{} with widely-used collective communication primitives. 
We also implement several benchmark applications that require the flexibility \thiswork, and achieve up to 3.99\texttimes{} speedup over the conventional inter-PE communication mechanism.
%
To the best of our knowledge, \scheme{} is the first framework to target fast and flexible inter-PE communication on commodity PIM-enabled DIMMs.
We will open-source \scheme{} once the paper gets accepted for publication.
We make the following key contributions: 
\begin{itemize}
    \item We reveal that commodity PIM-enabled DIMMs greatly suffer from high inter-PE communication overheads and limited applicability due to the lack of a fast and flexible communication framework.
    \item We propose \scheme{}, an inter-PE communication framework designed for the PIM-enabled DIMMs. \scheme{} abstracts the PEs as a virtual hypercube and allows collective communication along the chosen dimensions.
    \item \scheme{} supports eight widely used collective communication primitives and provides fast implementations optimized for PIM-enabled DIMMs. 
    \item \scheme{} greatly improves the performance of 
    applications involving diverse and complex inter-PE communication patterns with real commodity PIM-enabled DIMMs.
    \item We will publish \thiswork available as open-source, to facilitate research along similar directions.
\end{itemize}

\section{Background}
\label{sec:background}


\subsection{PIM-enabled DIMMs and Entangled Groups}
\label{sec:back:pid}


\begin{figure}
    \centering
    \includegraphics[width=\columnwidth]{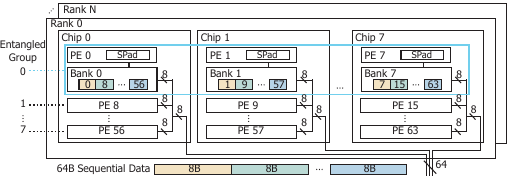}
    \caption{Internal architecture of commodity PIM-enabled DIMMs.}\vspace{-5mm}
    \label{fig:pid}
\end{figure}

PIM-enabled DIMMs implement PIM on dual in-line memory modules (DIMMs) with processing elements (PEs) attached near their memory banks~\cite{bufcmp, upmem, trim}.
\cref{fig:pid} depicts the internal hierarchy of such architecture~\cite{upmem}. 
Following modern DDR DIMM architectures~\cite{ddr4}, a channel is composed of multiple ranks that are independent but share the same 64-bit external bus.
A rank comprises multiple chips inside (usually 8).
In such a setting, each chip has 8-bit buses, which are concatenated to form the 64-bit bus of a channel.
Then, a chip contains multiple banks (usually 8).
The chips in a rank work in unison, so accessing bank 0 of chip 0 will also access bank 0 of chips 1--7 simultaneously.
Because of this, a 64b word is split into 8b segments and gets spread among the 8 banks (e.g., banks 0--7 or banks 56--63).
We refer to those sets of banks (or the associated PEs) as an \emph{entangled group}.
Accessing an entire entangled group together is critical for utilizing all the memory bandwidth from 64-bit bus.
Commodity PIM-enabled DIMMs attach a PE and a scratchpad memory, called working random access memory (WRAM), to each of their memory banks.
The memory banks are referred to as the main random access memories (MRAMs).
As the PEs can directly access the banks, the aggregate bandwidth from the PEs to the banks is an order of magnitude higher than that of the external bus.

\subsection{Domain Transfer}
\label{sec:back:dt}
Due to the aforementioned segmented data placement, a PE would not be able to process data elements larger than 8 bits.
To handle this, when data are transferred between the PIM-enabled DIMMs and host CPUs, the device driver~\cite{upmemsdk} running on the CPUs automatically rearranges the data behind the scenes using vector instructions so that a full 64b word (bytes 0-7 in \cref{fig:pid}) can be placed in a single bank.
We refer to this as \emph{domain transfer} between the PIM domain and host domain;
the data are not lost, but cannot be interpreted when placed as-is in the other domain.
Domain transfer rearranges the first 8 bits of each element in the first 8 bytes of the vector register, the second 8 bits in the next 8 bytes, and so on until all 64 bytes of sequential data are reordered.  
This ensures that once the 64-byte domain-transferred data is transferred to PEs, a full 8-byte element occupies each bank.
Although this process is transparent to the end users, this domain transfer incurs large performance overheads as revealed by prior work~\cite{prim, pidjoin, pimulator, upmemtraining, sparsep} and will be demonstrated in \cref{sec:motiv}.

\subsection{Collective Communications}
\label{sec:back:collcomm}

\begin{figure}
    \centering
    \includegraphics[width=\columnwidth]{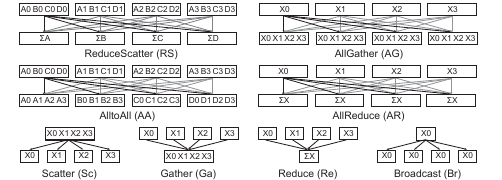}
    \caption{Illustrations of eight representative collective communication primitives among four nodes.}\vspace{-4mm}
    \label{fig:collcomm}
\end{figure}

Collective communication refers to a set of communication patterns widely used for parallel programming, where multiple nodes are coordinated to exchange data and perform computations. 
 It is widely used in high-performance computing~\cite{mpicharacterization, laguna2019largescale}, and more recently by machine learning~\cite{nccl, sccl, megatron, zero}.
Due to its popularity, collective communication libraries for CPUs~\cite{mpitharkur, chantheory}, graphics processing units (GPUs)~\cite{nccl, sccl}, and further optimizations for CPUs and GPUs have been proposed~\cite{themis, blueconnect, mscclang, flexreduce}.
However, these do not fit into PIM-enabled DIMMs because the libraries optimize against physical multi-hop networks.
On the other hand, the PIM-enabled DIMMs have no direct interconnection between banks, and the main challenge is to reduce the host CPUs' burden rather than minimize the network traversal costs.

We depict eight widely used~\cite{nccl, gloo, sccl} collective communication primitives in \cref{fig:collcomm}. 
ReduceScatter (RS), AllGather (AG), AlltoAll (AA), and AllReduce (AR) involve multiple PEs collaborating, each contributing to a segment of the data and then receiving a portion of the aggregated or modified data.
For example, in AllReduce, the four participating nodes have one element ($X_0-X_3$) each, and after communication, each node has the sum of all the elements ($\sum X$).
In contrast, Broadcast (Br), Reduce (Re), Scatter (Sc), and Gather (Ga) are done by direct host and PE communication. 
\thiswork supports all listed collective communications.

\section{Motivation} 
\label{sec:motiv}


\subsection{Conventional Communication Models and Libraries}
\label{sec:moti:conventional}
It is well known that commodity PIM-enabled DIMMs suffer from poor inter-PE communication performance~\cite{prim, pidjoin, pimulator}.
\cite{prim} classifies applications involving inter-PE communication as ``PIM-unfriendly'', where 
similar phenomenons are repeatedly reported in many literature~\cite{pidjoin, sparsep, upmemtraining}.

This is largely because the communication model has not been well-established nor optimized for PIM-enabled DIMMs.
Commodity UPMEM's SDK~\cite{upmemsdk} defines their programming model as a shared-nothing structure~\cite{upmem} with symmetric PEs.
This means that each PE is meant to run independently until the control returns to the host, and the PEs have no specific hierarchy among them.
While some programming models for PIM exist with (albeit limited) consideration on communication~\cite{bufcmp, nda, simplepim}, they all lack enough performance and flexibility due to the host processor bottleneck.

\begin{figure}[t]
    \centering
    \includegraphics[width=\columnwidth]{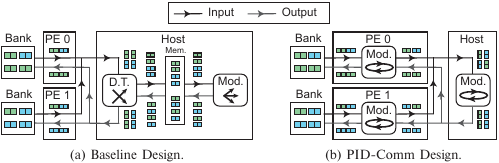}
    \caption{Communication flow of prior work and \thiswork.}\vspace{-3mm}
    \label{fig:moticomm}
\end{figure}

In \cref{fig:moticomm}(a), we illustrate how the lack of a well-defined inter-PE communication model results in a heavy host processor burden in AlltoAll with the conventional approach. 
The figure represents a toy example where two PEs within an entangled group share two 16-bit words (a small box represents a byte). 
The color represents the destination of each word: green is destined for PE 0, and blue is for PE 1.

First, the data are sent from PEs to the host, which goes through domain transfer (\cref{sec:back:dt}) and is put on the host memory.
Then, data modulation is performed, which contains globally re-arranging the data layout for AlltoAll.
Performing domain transfer again and putting the data back to the PEs concludes the communication.


\begin{figure}[t]
    \centering
    \includegraphics[width=\columnwidth]{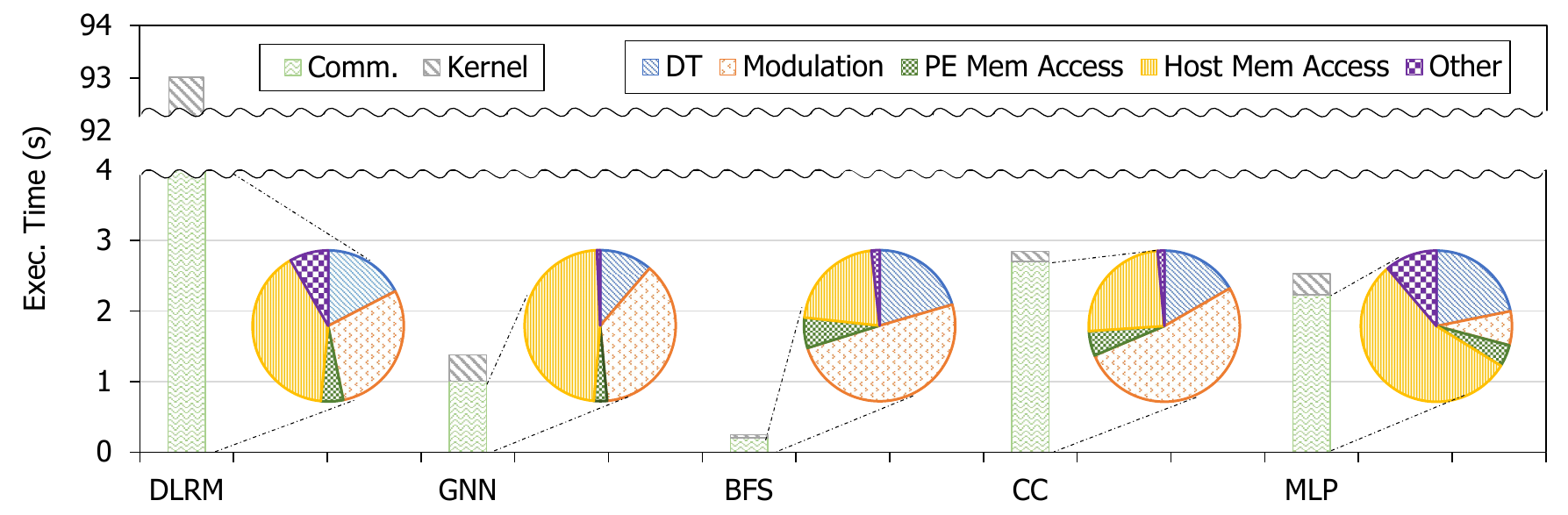}
    \caption{Execution time breakdown of applications on PIM-enabled DIMMs.}\vspace{-4mm}
    \label{fig:motiperf}
\end{figure}

Unfortunately, this model has three severe bottlenecks, as we exemplify in \cref{fig:motiperf} with breakdowns of benchmark applications (see \cref{sec:apps} for details).
In all five applications, the communications consume a substantial amount of time.
Further investigation into the communications (pie charts) reveals where the time is spent.
First, data modulation is done single-handedly by the host. 
Second, storing the data in host memory causes a large overhead. 
%
Third, domain transfer (DT) is done for all data that reaches the host. 

\cref{fig:moticomm}(b) shows an ideal flow where these bottlenecks are removed.
However, such an ideal flow is not straightforward to implement. 
For example, the host cannot interpret data without domain transfer, and the working set size well exceeds the size of the cache size, necessitating the use of host memory.
Despite the challenges, \thiswork addresses the issues, greatly lifting the host processor's burden and achieving high communication throughput.

\subsection{Lack of a Flexible Communication Model}
\label{sec:moti:multi}
\begin{table}
    \centering
     \caption{Comparison Against Conventional Approaches}
    \resizebox{0.95\columnwidth}{!}
    {
    \setlength{\tabcolsep}{3pt}
    \begin{tabular}{ccccccccccc}
     \toprule
      \multirowcell{2}{} & \multirowcell{2}{Multi-Instance \\ Communication} & \multirowcell{2}{Performance} & \multicolumn{8}{c}{Supported Primitives}\\
      \cline{4-11} &&&  \makecell{AA} &  \makecell{RS} &  \makecell{AG} &  \makecell{AR} & \makecell{Sc} & \makecell{Ga} &\makecell{Re}  &  \makecell{Br} \\
      \hline
      UPMEM SDK \cite{upmem} & Not Supported & Not Optimized &  &  &  &  & \cmark & \cmark &  &\cmark\\
      \hline
      SimplePIM \cite{simplepim} & Not Supported & Not Optimized &  &  & \cmark & \cmark & \cmark & \cmark &  & \cmark\\ 
      \hline
      \makecell{\textbf{PID-Comm} \\ \textbf{(Proposed)}} & \textbf{Supported} & \textbf{Optimized} & \cmark & \cmark & \cmark & \cmark & \cmark & \cmark & \cmark & \cmark\\
      \bottomrule\vspace{-7mm}
\end{tabular}
    }
       
    \label{tab:prior}
\end{table}



In addition to high-performance communication, there is a need for more flexible communication model for practical uses.
This is because 
many modern applications~\cite{dlrm, cagnet, fft, megatron-lm} require multiple instances of collective communications performed together across diverse dimensions of communication groups during their lifetime~\cite{mpicharacterization}.
Despite the need, as shown in \cref{tab:prior}, all the existing approaches~\cite{pidjoin, upmemtraining, simplepim} only allow a singular communication primitive on a fixed set of PEs. 
This confines the use of PIM-enabled DIMMs on a limited subset of applications.
To alleviate this issue, we need a flexible inter-PE communication model that can support interaction among subsets of PEs in several dimensions.

At the same time, allowing communications between an arbitrary subset of PEs could lead to a slowdown. 
For example, \cref{fig:pid} shows that it requires at least 8 PEs in an entangled group to draw maximum transfer bandwidth from the DIMMs.
If communication is performed between 8 PEs from the same chip instead, it will greatly suffer from low bandwidth.
Hence, a good communication model should guide the user such that efficient communication can be achieved without much effort.

\section{\thiswork Communication Model}
\label{sec:model}

\subsection{Design Goals and Challenges}
\label{sec:model:goals}
In this section, we describe our design for the proposed communication model.
Following the motivation from \cref{sec:motiv}, we aim to achieve two goals. 
First, the model should provide enough \emph{flexibility} for users to support diverse dimensions in various applications. 
Second, the model should lead to \emph{high-performance} implementation, regardless of the user's configuration for the communication instances.

Existing communication libraries for CPUs~\cite{mpitharkur} and GPUs~\cite{nccl} typically use the abstraction of \emph{communication group}, which is essentially a user-defined subset of the workers. 
For multi-instance communications, users would define multiple communication groups and initiate the desired communication for each group.
While using the same abstraction for the PIM-enabled DIMMs will allow multi-instance communications, allowing any arbitrary subset of PEs to form communication groups could sabotage the performance. 

\begin{figure}
    \centering
    \setlength{\fboxsep}{0.5pt}
    \setlength{\fboxrule}{1pt}
    \includegraphics[width=\columnwidth]{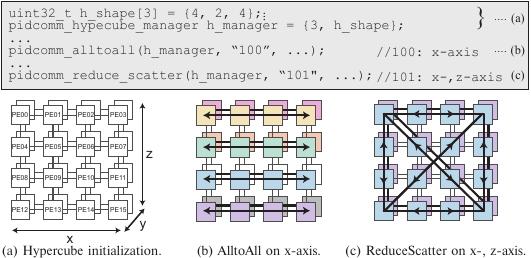}
    \caption{Virtual hypercube and multi-axis communication topology.}\vspace{-5mm}
    \label{fig:hypercube}
\end{figure}

\subsection{Virtual Hypercube Communication Model}
\label{sec:model:hypercube}
To achieve both goals, we propose a communication abstraction model named \emph{virtual hypercube}, which includes three design choices listed below.
\subsubsection{User-defined hypercube configuration} For an application, users define a hypercube with an arbitrary number of dimensions whose number of nodes matches the number of used PEs. 
The length of each dimension is also chosen by the user as a power-of-two integer, except for the last dimension. 
For example, \cref{fig:hypercube}(a) shows an example 4$\times$2$\times$4 hypercube for 32 PEs. 
The nodes of the virtual hypercube are transparently mapped to the physical PEs by the \thiswork library. 

\subsubsection{Cube slices as communication groups} For each communication instance, multiple communication groups are defined at once by selecting the dimensions that compose each group.
\cref{fig:hypercube} shows two examples, where the (b) contains $x$ dimensions for the group, which leads to 4$\times$2 communication groups of size 4. 
(c) choose $x$ and $z$ dimensions, which results in 2 communication groups of size 16.

\subsubsection{Multi-instance invocation} On the sliced cubes, multiple instances of collective communications are invoked together. For communication primitives without a root (e.g., AlltoAll), invoking a communication primitive will start the communication on all cube slices.
For example, on \cref{fig:hypercube}(b), eight AlltoAll instances are conducted independently of each other. 
For the primitives with a root (e.g., Reduce), the host processor always becomes the root. 
The communication groups independently communicate with the host, where the host has separate buffers assigned to communicate with each group.


This hypercube abstraction guides users through the constraints for drawing the maximum transfer bandwidth from the DRAM hierarchy. 
No matter how the groups are formed, this ensures that an entangled group can always operate as a whole to guarantee the maximum transfer bandwidth. 

\begin{figure}
    \centering
    \includegraphics[width=\columnwidth]{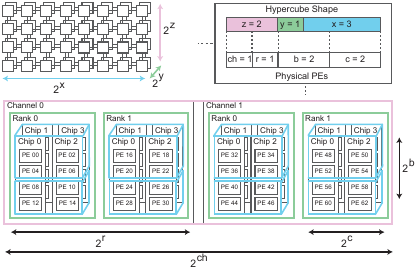} 
    \caption{Hypercube to physical PIM-enabled DIMM mapping.}\vspace{-4mm} 
    \label{fig:mapping}
\end{figure}

\subsection{Mapping Virtual Hypercube to Physical PEs}
\label{sec:model:mapping}
Because both the virtual hypercube and the DRAM hierarchy are regular, mapping between them is straightforward.
%
 First, we identify the entangled groups and use them as assignment units.
The existing frameworks~\cite{prim, simplepim} regard all PEs as symmetric and assume no hierarchy.
While such abstraction makes programming simple, this could cause a severe throughput drop as they ignore the underlying hierarchy, especially the entangled groups.
Identifying them and exposing them is the first step toward hypercube-to-PE mapping.

Second, the entangled groups fill the hypercube in order.
While there could be various valid mapping methods to do so, 
we follow the DRAM hierarchy in the order of \underline{c}hip-\underline{b}ank (PE)-\underline{r}ank-\underline{ch}annel (entangled groups correspond to the `chip' hierarchy).
Because the only possible level with non-power-of-two length is the number of channels, placing this as the last dimension suits the order of the entangled groups.
A toy example mapping between $[ z=2, y=1,x=3]$ and $[ch=1, r=1, b=2, c=2]$ is illustrated in \cref{fig:mapping}. The $x=3$ dimension occupies two entangled groups of 4 chips. 
Then two such banks form $y=1$ dimension, while the channel and the rank fill the last dimension of $z=2$.

\begin{figure*}
    \centering
    \includegraphics[width=\textwidth]
    {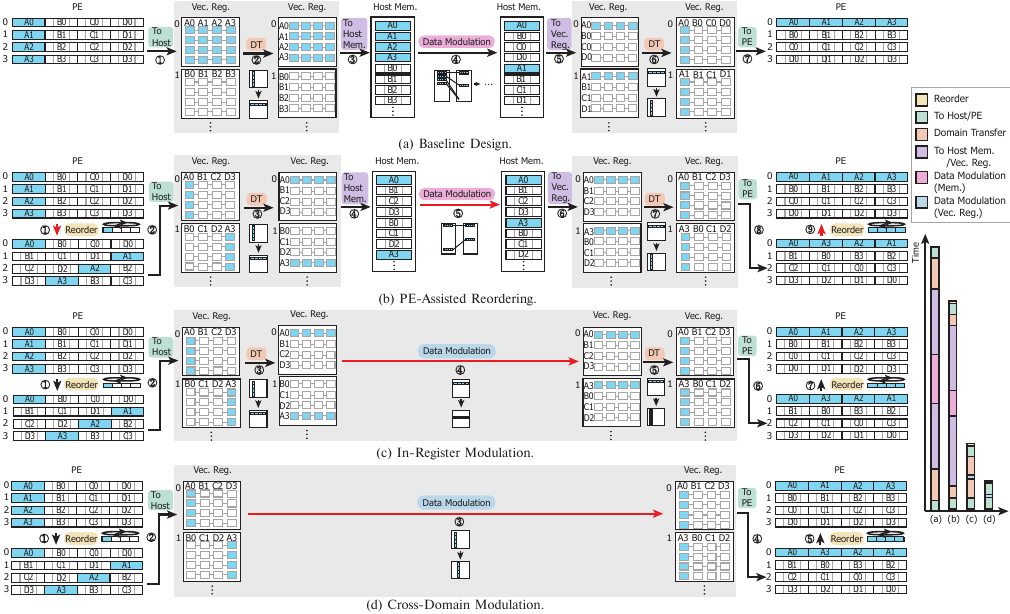}
    \caption{Proposed collective communication improvement techniques demonstrated with AlltoAll.}\vspace{-2mm}
    \label{fig:commlib}
\end{figure*}

\section{\thiswork Library} 
\label{sec:library}


To realize \thiswork's communication model, we design a new high-performance collective communication library for PIMM-enabled DIMMs.
%
Leveraging insights from 
the DRAM and PIM architectures, we introduce three novel techniques to alleviate the communication bottleneck. 
We describe the details of 
each technique 
using AlltoAll.
We then apply them to other collective communication primitives.

For simplicity's sake, the illustrations of the proposed design (\cref{fig:commlib} - \cref{fig:multiaxes}) assume entangled groups comprising 4 PEs, 32bit elements, and 128-bit vector registers.
In practice, more realistic sizes would be 8 PEs comprising an entangled group, 64-bit elements, and 512-bit vector registers. 
The diagrams can be naturally extended into such an environment. 
%



\subsection{\thiswork Performance Optimization Techniques}
\label{sec:design}

In \cref{fig:commlib}(a) the baseline design from \cref{fig:moticomm}(a) is illustrated in detail, following seven steps (\circled{1}-\circled{7}).
The words destined for PE0 ($A0-A3$) are colored \textcolor{blue}{blue} for easy tracking.
For demonstrative purposes, we take three progressive steps from the baseline to highlight the optimizations we apply.


\subsubsection{\Pr}
\label{sec:design:pr}
The initial challenge we address 
with \emph{PE-assisted reordering} 
is the execution burden of global data modulation (rearrangement) on the host. 
This is represented by data reordering on host memory in \cref{fig:commlib}(a) \circled{4} which distributes words (e.g., $A0-A3$) to four different addresses of the memory.

Our key observation is that the global rearrangement can be decomposed into three local rearrangements, two of which can be processed by the PEs in parallel.
With \pr, 
PEs perform local reordering before sending data to the host \circled{1} and once more after the data return \circled{9}.
This is done by uploading a part of data to WRAM, incrementally shifting it such that later each vector register can hold data with different destinations, and than rewrite the updated data back to MRAM within each PE.
The host still has to perform modulation, but the movements are now local, requiring fewer host instructions and also becoming cache-friendly.
This improves overall latency since the steps \circled{1} and \circled{9} are performed in parallel with multiple PEs. 
Note that this first design serves as an important cornerstone of subsequent techniques, each addressing the most significant bottlenecks depending on the specific communication primitive.

\subsubsection{\Im}
\label{sec:design:im}
Next, we remove 
host memory access. 
Due to \pr, the working set size of the modulation inside the host now fits a single vector register as in \circled{5} of \cref{fig:commlib}(b) (a word does not move outside the register size boundary).
Thus, we perform the host-side modulation only within the vector registers by carefully managing the data in a streaming fashion.
For AlltoAll, word-level shifts are performed with SIMD instructions.
Because this eliminates storing data in the host memory, this saves a great amount of time 
and host memory requirements.

\subsubsection{\Cm}
\label{sec:design:cm}
Finally, we remove domain transfers for primitives that do not use arithmetic operations.
Domain transfer is only needed when the words are processed in arithmetic operations in the host (i.e., reduction).  
Thus, domain transfers can be avoided for communication primitives that only redistribute data, such as AlltoAll and AllGather. 
Once \im is applied, the remaining host-side operations are domain transfer~\circled{3}, word-level shift~\circled{4}, and another domain transfer~\circled{5} as shown in \cref{fig:commlib}(c).
By fusing those three steps, they become a single byte-level shift operation (\circled{3} of \cref{fig:commlib}(d)).
This again saves a great amount of host-side instructions, leading to a huge speedup.


\begin{figure*}
    \centering
    \includegraphics[width=\textwidth]{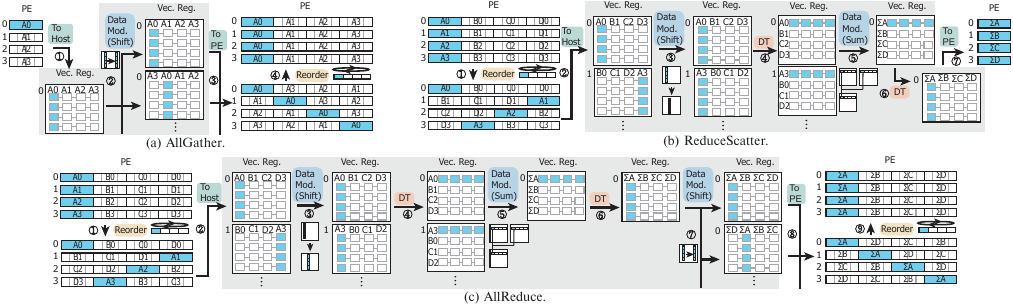}
    \caption{Proposed collective communication for (a) AllGather, (b) ReduceScatter, and (c) AllReduce.}\vspace{-4mm}
    \label{fig:commlibprim}
\end{figure*}

\subsection{Other Collective Primitives}
\label{sec:design:others}

\begin{table}
    \centering
     \caption{Applicability of the proposed techniques}
    \resizebox{.9\columnwidth}{!}
    {
    \setlength{\tabcolsep}{5pt}
    \begin{tabular}{lcccccccc}
           \toprule
      & AA & RS & AR & AG & Sc & Ga & Re & Br\\
            \midrule
      PIM-assisted reordering & \cmark & \cmark & \cmark & \cmark & & & \cmark &\\
      \midrule
      In-register modulation & \cmark & \cmark & \cmark & \cmark & \cmark & \cmark & \cmark &\\
      \midrule
      Cross-domain modulation & \cmark &  &  & \cmark &  &  &  &\\
      \bottomrule\vspace{-7mm}
    \end{tabular}
    }
       
    \label{tab:techniques}
\end{table}

We now describe how the techniques discussed in the previous subsection are implemented for communication primitives other than AlltoAll (AA). 
Also see \cref{tab:techniques} for which technique is applied to each primitive.

\subsubsection{AllGather (AG)}
AllGather closely resembles AlltoAll's procedure. 
Both do not perform arithmetic operations on the host, allowing all techniques from \cref{sec:design}, including \cm to be applied. 
The difference is that AllGather starts from each node having a data element, and all the nodes will hold the concatenated elements as a result. 

As in \cref{fig:commlibprim}(a), the data are first directly loaded into vector registers (\circled{1}). 
To send different elements to each PE, the host repeats the process of byte-level shift (\circled{2}, \cm and \im) and sends them (\circled{3}) back to the PEs. 
After this, each PE contains a copy of the complete values 
in different orders.
Fixing this misalignment 
completes AllGather (\circled{4}, \pr).

\subsubsection{ReduceScatter (RS)}
\label{sec:design:others:RS}
Unlike AlltoAll and AllGather, ReduceScatter requires arithmetic reductions on the host. 
Thus, domain transfer is necessary and \cm cannot be used. 
The first few steps of \cref{fig:commlibprim}(b) are the same as the steps \circled{1}-\circled{3} from those of AlltoAll in \cref{fig:commlib}(d) in that it reorders in PEs (\pr) and performs byte-level shift.
Afterward, the host performs domain transfer on the data (\circled{4}), 
followed by the host performing vector addition using SIMD instructions to obtain a vector register containing all reduced values (\circled{5}, \im). 
This is possible since PEs arrange the data such that the elements to be added together are located in different vector registers but in the same slots. 
This approach is essential in that vertical reduction may be done with a single SIMD instruction per vector register whereas horizontal in-register reductions require multiple costly operations.
Finally, performing domain transfer (\circled{6}) and sending to the PEs (\circled{7}) completes ReduceScatter.

\subsubsection{AllReduce (AR)}
In many collective communication libraries~\cite{mpitharkur, nccl, mscclang, blueconnect, flexreduce}, AllReduce is implemented as a naive combination of ReduceScatter and AllGather. 
These libraries benefit from direct communication links between nodes, eliminating the need to reroute to other devices.
However, in PIM-enabled DIMMs, using 
the same strategy would only involve multiplied external bus usage. 

Instead, we define a separate AllReduce primitive by seamlessly merging ReduceScatter and AllGather as in \cref{fig:commlibprim}(c).
After following the steps of \circled{1}-\circled{6} same as ReduceScatter, all data are reduced inside the host memory and domain transferred utilizing \pr and \im.
Sending the reduced data to PEs is done similarly to steps \circled{2}-\circled{4} of AllGather (\circled{7}-\circled{9} of \cref{fig:commlibprim}(c)). 

\subsubsection{Primitives with Roots}
With the choice of always having the host as the root, the collective communication primitives with roots can be implemented using partial routines of the primitives from previous subsections.
Splitting ReduceScatter into half, 
\circled{1}-\circled{5} becomes Reduce (Re), 
and \circled{6}-\circled{7} becomes Scatter (Sc).
From AllGather, step \circled{1} followed by domain transfer gives Gather (Ga).
Step \circled{3} preceded by domain transfer gives Broadcast (Br), which is equal to the broadcast implementation from \cite{upmemsdk}.

\begin{figure*}
    \centering
    \setlength{\fboxsep}{0.5pt}
    \setlength{\fboxrule}{1pt}
    \includegraphics[width=\textwidth]{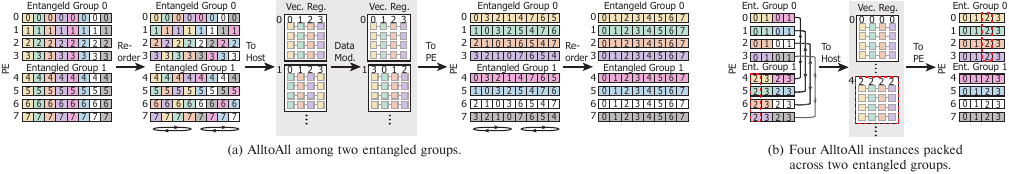}
    \caption{AlltoAll communication being performed on two general cases.}\vspace{-4mm}
    \label{fig:multiaxes}
\end{figure*}

\subsection{Extension to General Cases}
\label{sec:general}

Previously, we explained the design of \thiswork using the case of a single entangled group of 64b words.
We next discuss issues of generalizing the above implementations. 

\emph{Data types.} 
Data types smaller than 64b are supported with a straightforward extension. 
For example, with 32b items, one can assume the same dataflow, but two 32b items are packed in a single 64b chunk.
Only in the primitives with arithmetic operations should the data be regarded as lower types instead of 64b items.
One interesting exception is 8b items, because the host CPU can interpret and calculate data without the domain transfer. 
Thus, domain transfer from ReduceScatter and AllReduce is removed accordingly.

\emph{Larger dimensions.} 
When the number of participating nodes per dimension is larger than eight, we have more than two entangled groups involved together.
For example, \cref{fig:multiaxes}(a) shows a case where AlltoAll is performed among two entangled groups.
This is done by naturally extending the principles proposed in \cref{sec:design}.
For \pr, the data within each PE is partitioned into two (the number of entangled groups) and rotated individually.
Other techniques can be similarly applied, and communications among more than two entangled groups are processed the same way.

\emph{Multi-instance communications over different dimensions.}  
With the multi-instance communications using hypercube abstraction, the communication groups might be formed across several entangled groups as in \cref{fig:multiaxes}(b).
This seemingly complicated problem is not very difficult to solve, as it only involves loading the data into a vector register and rewriting them to the PIM in different addresses.
For example, the red dotted box comes from the first slots of entangled group 1 (PEs 4--7) into a single vector register, which is written unmodified to the third slots of the entangled group 0 (PEs 0--3).
This is possible because 
PEs within an entangled group already hold words with different destination PEs. 
Additionally, the user's hypercube configuration might split the entangled group into half or quarters, which is handled similarly.





\section{\thiswork Programming Framework}
\subsection{Communication Framework}

\begin{figure}
    \centering
    \setlength{\fboxsep}{0.5pt}
    \setlength{\fboxrule}{1pt}
    \includegraphics[width=\columnwidth]{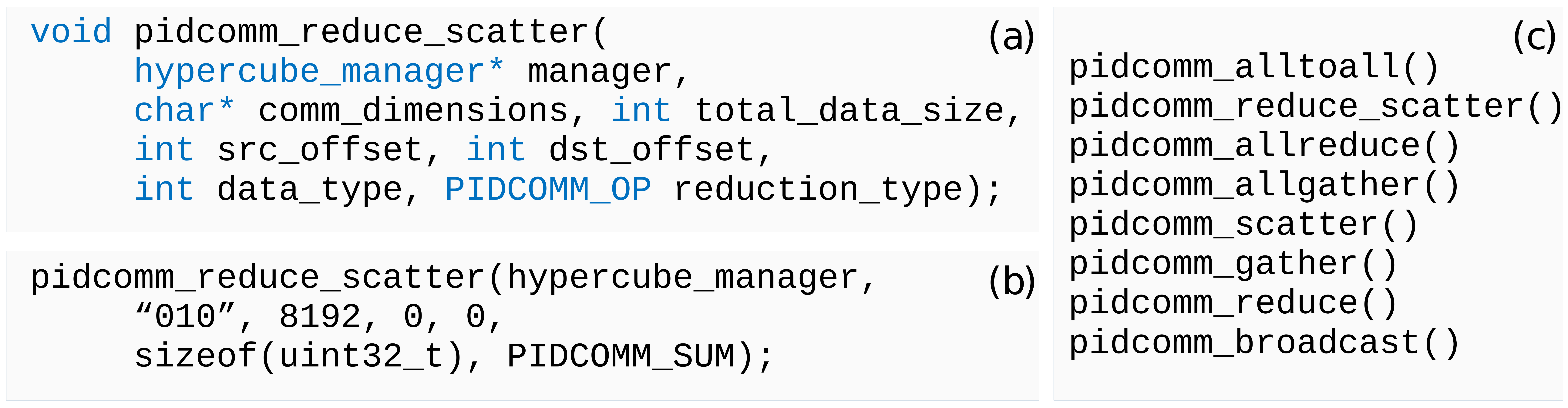}
    \caption{PID-Comm APIs. (a) lists arguments for ReduceScatter, (b) shows how to use it. (c) lists the supported primitives.}\vspace{-4mm}
    \label{fig:pidcomm_apis}
\end{figure}
\thiswork provides communication primitives through various APIs 
as the host-side code.
For end-users' convenience, PID-Comm's APIs are similar to popular communication frameworks such as MPI~\cite{mpitharkur} or Nccl~\cite{nccl}.
Unlike other distributed frameworks, \thiswork requires additional variables as shown in \cref{fig:pidcomm_apis}(a): {\textit{pidcomm\_hypercube\_manager}} for configuring a hypercube, and {\textit{comm\_dimensions}} represented by a bitmap string that contain the target dimensions.

An example of performing ReduceScatter along the y-axis in a 3D hypercube is shown in the \cref{fig:pidcomm_apis}(b). Note that the {\textit{comm\_dimensions}} ``y-axis'' is represented as {\textit{``010''}}. 
\cref{fig:pidcomm_apis}(c) lists the eight APIs supported by \thiswork.

\cref{algorithm_gnn} shows how this can be used to implement an application in pseudo-code. 
Between PIM kernel calls, communication primitives are inserted with user-defined dimensions. 
Note that each communication invocation is worth around few hundred LOC if implemented in an ad-hoc manner.


\begin{algorithm} \caption{An Example code with PID-Comm (GNN)} \label{algorithm_gnn}
\begin{algorithmic}[1]
\State Initialize hypercube\_manager (2D);
\State pidcomm\_scatter();\Comment{Send initial data to PEs}
\Repeat
\State dim = ``01''  $\rightleftarrows$ ``10''\Comment{Alternating dimension} 
\State PE\_kernel(SpGEMM);
\State pidcomm\_reduce\_scatter(dim);\Comment{Features}
\State PE\_kernel(GeMM);
\State pidcomm\_allreduce(dim);\Comment{Features}
\Until {All layers are processed}
\State pidcomm\_reduce();\Comment{Achieve final results}
\end{algorithmic}
\end{algorithm}\vspace{-1mm}

\subsection{Implementation}
\label{sec:impl}
We implement \thiswork on top of UPMEM SDK~\cite{upmemsdk}.
We manipulated the conventional library to disable automatic domain transfer, allowing PIM-domain redistribution inside the host.
We also reverse-engineered the default mechanism of allocating kernels to PEs to design a new method of using the entangled groups within the DIMM hardware hierarchy. 
This enabled the implementation of the hypercube abstraction and mapping.
All host-side computations are done using AVX-512 instructions~\cite{avx512}, such as {\textit{\_mm512\_rol\_epi64()}} for byte level shifts.
Additionally, the operations are conducted in 
DDR4 burst granularity (i.e., 64 bytes) to fully utilize the external bus. 
With these techniques, \thiswork's APIs provide: 
synchronizing PIM kernels to ensure the data are ready, invoking the PIM-side preparation kernels (i.e., reordering), data movement, executing the host-side code, and invoking the PIM-side post-processing kernel.
This frees the users from worrying about PIM-CPU split or synchronization.

\section{Benchmark Applications}
\label{sec:apps}

To demonstrate how \thiswork can be used to implement important parallel applications, we implement five applications listed in \cref{tab:baselines}: Deep learning recommendation model (DLRM), graph neural networks (GNN), breadth-first search (BFS), connected components (CC), and multi-layer perceptron (MLP). 
For applications where reference PIM implementations exist (GNN, BFS, MLP), we faithfully validated results against them, or against CPU-based ones otherwise (DLRM, CC).
For all applications, we optimized PIM baselines and proposed kernels to utilize the scratchpad memory and spawn enough tasklets.
These kernels were launched on the PEs to accelerate benchmark-specific computations.
Each application starts with Scatter to spread data and retrieves results to the host using Gather or Reduce.
For the CPU-only kernels in \cref{sec:eval:cpu}, we used benchmarks BFS and MLP from \cite{prim} and GNN from \cite{sparsep}. The remaining benchmarks were implemented using similar techniques from \cite{prim}.

\newcommand{\mytriangle}[1]{\tikz{\filldraw[draw=#1,fill=#1,color=orange] (0,0) --
(0.6em,0) -- (0.3em,0.6em);}}
\begin{table}[b]\vspace{-2mm}
    \centering
     \caption{Benchmark Applications}
    \resizebox{0.95\columnwidth}{!}
    {
    \setlength{\tabcolsep}{3pt}
    \begin{tabular}{cccccccccccc}
    \toprule
        \multirowcell{2}{App.}&\multirowcell{2}{Hyper.\\Dim.} &\multicolumn{8}{c}{Communication Primitives} &\multirowcell{2}{Datasets} &\multirowcell{2}{Environment}\\
        \cline{3-10}
         &&  SC & {GA} &{RD}  &  {BC} &  {AA} &  {RS} &  {AG} &  {AR}\\
        \midrule
        DLRM  & 3 & \cmark & \cmark &   & \cmark & \cmark & \cmark & &  & Criteo~\cite{criteo} & Emb. dim = 16, 32\\
        \midrule
        GNN RS\&AR & 2 & \cmark &  & \cmark   &  &  & \cmark &  & \cmark & PM~\cite{pubmed} RD~\cite{reddit} & Layers = 3\\
        GNN AR\&AG  & 2 & \cmark & \cmark &    &  &  &  & \cmark & \cmark & PM~\cite{pubmed} RD~\cite{reddit}  & Layers = 3\\
        \midrule
        BFS &  1 & \cmark &  & \cmark   & \cmark &  &  &  & \cmark & LJ~\cite{livejournal} LG~\cite{gowalla} & \\
        \midrule
        CC &  1 & \cmark &  & \cmark   & \cmark &  &  &  & \cmark & LJ~\cite{livejournal} LG~\cite{gowalla} & \\
        \midrule
        MLP  & 1 & \cmark &  & \cmark   &  &  & \cmark &  &  &   \multicolumn{2}{c}{Features = 16k, 32k, Layers = 5}\\

        \bottomrule\vspace{-7mm}
    \end{tabular}
    }
       
    \label{tab:baselines}
\end{table}

\begin{figure}
    \centering
    \setlength{\fboxsep}{0.8pt}
    \setlength{\fboxrule}{1pt}
    \includegraphics[width=\columnwidth]{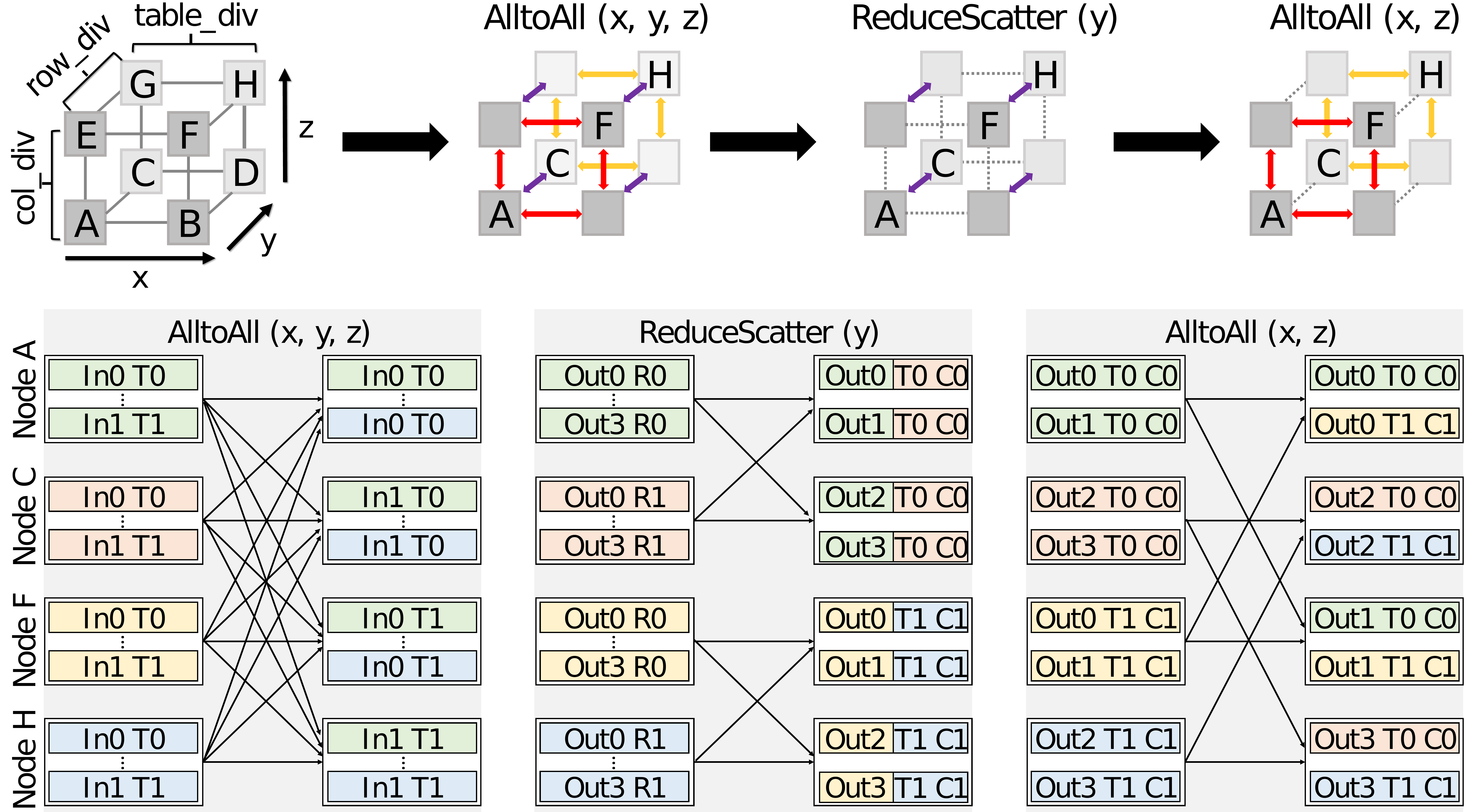}
    \caption{DLRM Communication Structure}\vspace{-4mm}
    \label{fig:dlrm}
\end{figure}

\subsection{Deep Learning Recommendation Model}
Deep Learning Recommendation Model~\cite{dlrm} (DLRM) is widely used to recommend products or advertisements with user information. 
We implemented DLRM from scratch and validated it on the reference implementation~\cite{dlrm}.
We use the Criteo dataset~\cite{criteo}, with embedding dimensions 16 and 32.
The DLRM architecture is split into three dimensions~\cite{dlrmisca} into tables, rows, and columns, which we map into the 3D hypercube. 
\cref{fig:dlrm} shows the mapping with how four chosen PEs (A, C, F, and H) communicate with each other.
The input batch is split by the divided embedding tables using an AlltoAll in the xyz-space (all PEs).
After the lookup, the complete embedding vector is obtained by performing ReduceScatter along the y-axis due to the row-wise parallelism.
Finally, AlltoAll for the xz-plane is performed to relocate embedding vectors in the correct PEs for the remaining linear layers.



\begin{figure}
    \centering
    \includegraphics[width=\columnwidth,trim={0 0 0 0},clip]{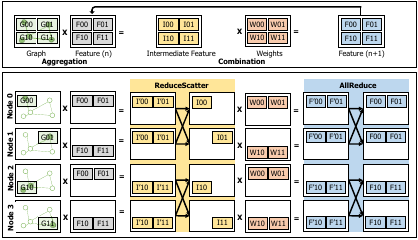}
    \caption{GNN (RS\&AR) Communication Structure. }\vspace{-4mm}
    \label{fig:gnn}
\end{figure}

\begin{figure*}
    \centering
    \setlength{\fboxsep}{0.5pt}
    \setlength{\fboxrule}{1pt}
    \includegraphics[width=0.88\textwidth]{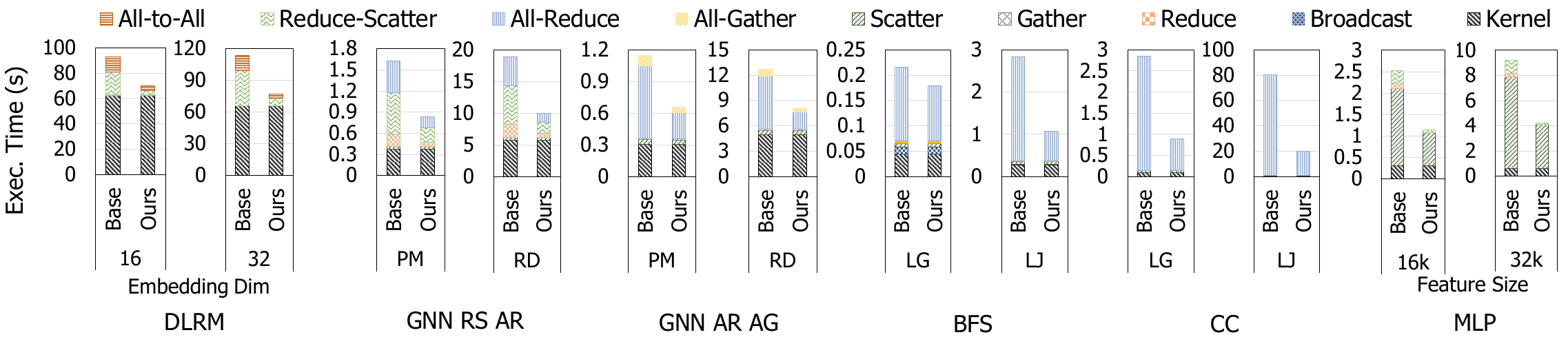}
    \caption{Performance of \thiswork on benchmark applications.}\vspace{-4mm}
    \label{fig:eval:benchmark}
\end{figure*}

\subsection{Graph Neural Networks}
Graph Neural Network (GNN) is a 
deep neural network that takes a graph as input. 
A GNN layer includes aggregation (SpGEMM) and combination (GeMM).
For the SpGEMM, we started from the state-of-the-art SpMV implementation SparseP~\cite{sparsep} and partitioned the feature matrix 
to fit into the scratchpad memory. 
We implemented GeMM from scratch.

We devise two 2D strategies similar to \cite{cagnet, snf}.
In the RS\&AR version in \cref{fig:gnn} and \cref{algorithm_gnn}, PEs process a 2D tile of the graph topology and a horizontal strip of tiles in the features. 
After aggregation, each PE has a partial sum of horizontal strips. 
They are ReduceScatter'ed and multiplied to weights for combination.
This will again result in partial sums, which are AllReduce'ed for aggregation in the next layer.
The AR\&AG version applies AllReduce after aggregation, whose combination leads to 2D tiled results. 
AllGather then prepares the features in horizontal strips for the next layer.


%

\subsection{Breadth First Search}
Breadth-first search (BFS) searches for the shortest path to a desired node in a given graph. 
In each iteration, the kernel traverses the list of visited nodes, and the neighbors of visited nodes are marked as visited until there are no unvisited neighbors. 
%
We used the reference implementation~\cite{prim} with some optimizations as our baseline. 
In each iteration, AllReduce is used to update the visited list of each node with $or$ reduction. 

\begin{figure}
    \centering
    \setlength{\fboxsep}{0.5pt}
    \setlength{\fboxrule}{1pt}
    \includegraphics[width=0.88\columnwidth]{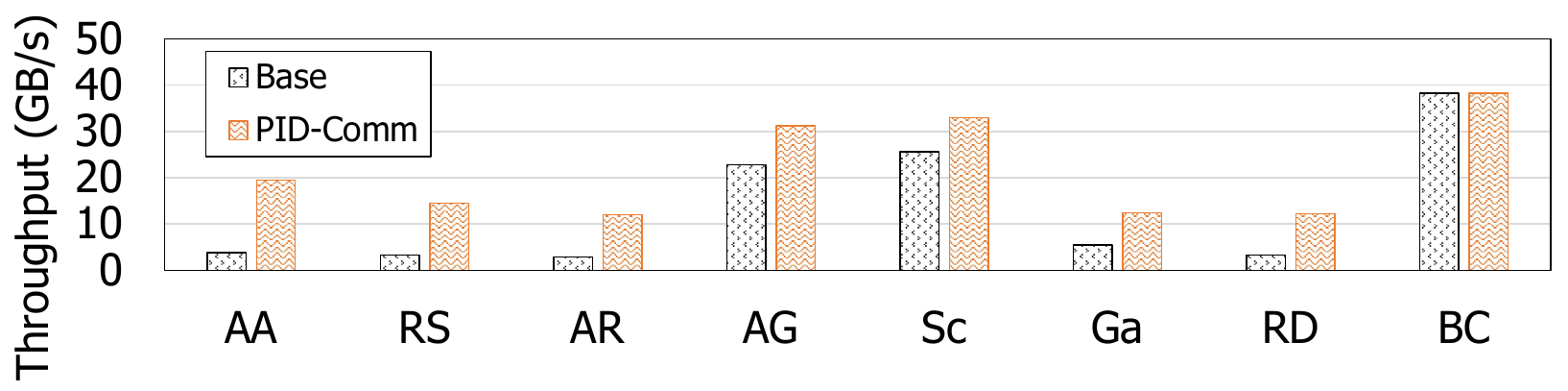}
    \caption{Performance of supported primitives.}\vspace{-3mm}
    \label{fig:eval:centprim}
\end{figure}

\begin{figure}
    \centering
    \setlength{\fboxsep}{0.5pt}
    \setlength{\fboxrule}{1pt}
    \includegraphics[width=0.88\columnwidth]{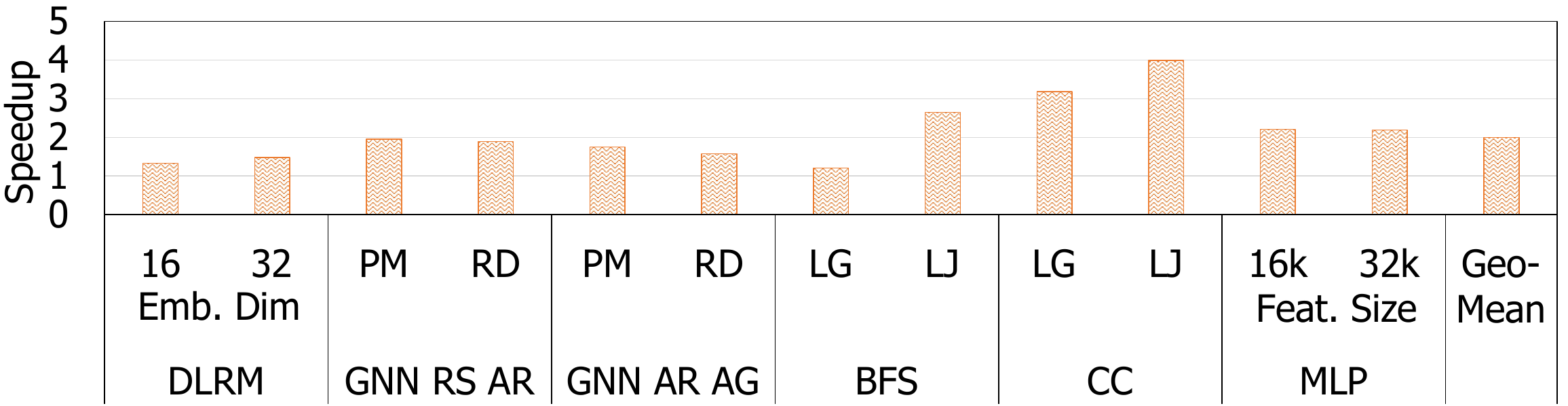}
    \caption{Speedup of benchmark applications.}\vspace{-5mm}
    \label{fig:eval:speedup_bench}
\end{figure}

\subsection{Connected Component}
Connected Component (CC) is another graph processing application used to search for the number of components in a given graph. 
CC utilizes a similar communication structure from BFS, which uses AllReduce for each iteration with $min$ reduction. 
We preprocessed directed edges to undirected edges. 

\subsection{Multi-layer Perceptron}
Multi-layer Perceptron (MLP) is a feedforward 
neural network widely used in machine learning. 
We used the reference implementation~\cite{prim} with 16k\texttimes16k and 32k\texttimes32k weight parameters.
We applied minor communication structure optimizations 
by column-wise partitioning the feature matrix. 
After a layer is computed, partial results from PEs are ReduceScatter'ed to each PE before proceeding to the next layer. 

\section{Evaluation}
\label{sec:eval}

\subsection{Experimental Setup}
\label{sec:eval:setup}
We have implemented and evaluated \thiswork on an UPMEM~\cite{upmem}-enabled system. 
The system is equipped with an Intel Xeon Gold 5215 CPU, and four channels of PIM-enabled DIMMs, each with four ranks.
This accumulates to 4 (Channels)\texttimes 4 (Ranks)\texttimes 8 (Chips)\texttimes 8 (Banks) = 1024 PEs.
For baselines, we used implementations from SimplePIM~\cite{simplepim} for the supported primitives. 
We faithfully implemented and optimized other non-supported primitives (AlltoAll, ReduceScatter, and Reduce) following the same principles. 
In addition, we implemented a multi-dimensional hypercube on top of the baseline for fair comparisons.
Note that all evaluations of baselines and \thiswork are performed on real-world systems attached with PIM-enabled DIMMs.

\subsection{Performance of Supported Primitives}
\label{sec:eval:collcomm}


In \cref{fig:eval:centprim}, we compare the throughput obtained for each primitive.
We used the 2D configuration of (32,32) for both the baseline and \thiswork, where the throughput is defined as the larger side of the data size (i.e., before reduction) divided by the execution time.
\thiswork shows significantly improved throughput compared to the baseline design, especially for primitives with slow baselines.
AlltoAll, ReduceScatter, and AllReduce show 
5.19\texttimes, 4.46\texttimes, and 4.23\texttimes{} throughput improvements. 
The geomean improvement of all primitives was 2.83\texttimes.
The only exception is Broadcast, which already shows high enough throughput from native UPMEM driver~\cite{upmemsdk}. 
This is because Broadcast requires domain transfer only once for an element, which can be used for all the PEs, where it already reaches close-to-peak DRAM bandwidth.

\subsection{Performance of Benchmark Applications}
\label{sec:eval:benchmark}

In \cref{fig:eval:speedup_bench}, we show that using \thiswork for benchmark applications can significantly improve performance. 
All applications exhibited better performance compared to the baseline with speedups ranging from 1.20\texttimes{} to 3.99\texttimes{}. 
The geomean speedup of the benchmark applications is 1.99\texttimes{}.

This large speedup can be explained by breaking down the execution times into the eight supported primitives and computation (`Kernel') at \cref{fig:eval:benchmark}.
While Gather and Broadcast are included in the figure, the percentage is at max 7\%. causing both to have a small impact on the overall execution time.
The communication latency for all applications is largely reduced. 
Thus, the high speedup is observed for applications with a larger communication portion (e.g., CC), and relatively lower speedup for applications with a smaller portion (e.g., DLRM).

\subsection{Ablation Study} 
\label{sec:eval:ablation}

\begin{figure}
    \centering
    \setlength{\fboxsep}{0.5pt}
    \setlength{\fboxrule}{1pt}
    \includegraphics[width=0.88\columnwidth]{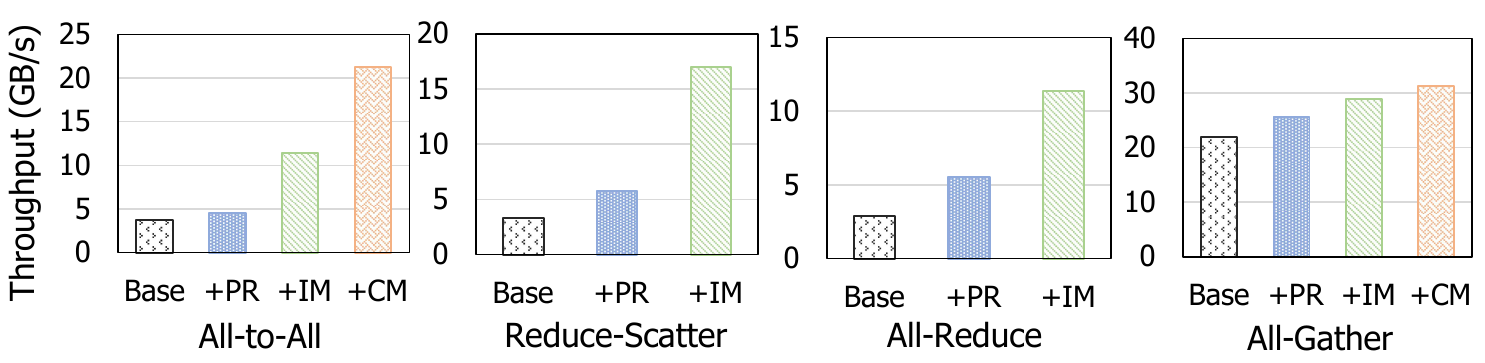}
    \caption{Ablation study of \thiswork primitives.}\vspace{-4mm}
    \label{fig:eval:ablation}
\end{figure}

In \cref{fig:eval:ablation} we measured the effectiveness of each inter-PE communication design through an ablation study of 
four inter-PE primitives: AlltoAll, ReduceScatter, AllReduce, and AllGather.
Overall, the results show a consistent trend of improved throughput as new techniques are applied. 
Our first technique is \pr, which increases the throughput by 1.48\texttimes{} in geomean and is the most effective in ReduceScatter and AllReduce. 
This is caused by host reduction being more computation-intensive than data reordering.
Thus, PEs aiding part of the host's operation had a higher impact than other non-arithmetic primitives.
Next, applying \im additionally increases the throughput 2.03\texttimes{} in geomean. 
For ReduceScatter and AllReduce, this technique is the most effective by removing the host memory access.
Finally, for non-arithmetic primitives such as AlltoAll and AllGather, applying \cm achieves 1.42\texttimes{} throughput improvement in geomean. 
Note that the overall intensity of throughput improvement was relatively smaller for AllGather, as the baseline was already fast enough.


\begin{figure}
    \centering
    \includegraphics[width=0.86\columnwidth]{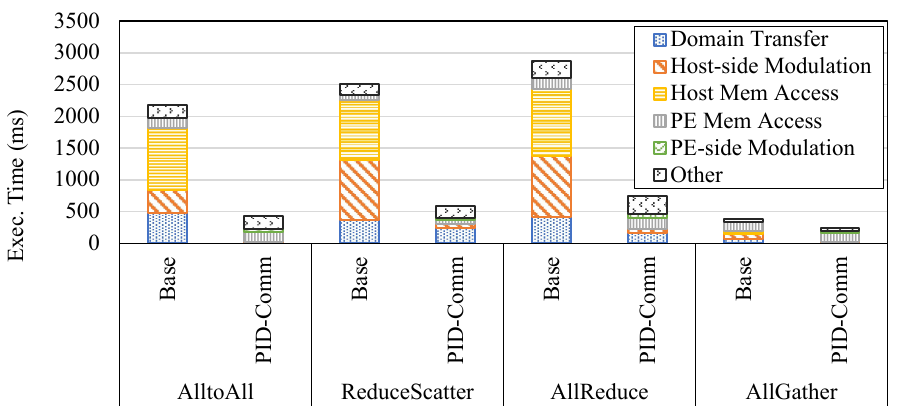}
    \caption{Breakdown of \thiswork primitives.}\vspace{-3mm}
    \label{fig:eval:Primitive_breakdown}
\end{figure}

We further analyze how each inter-PE communication optimization technique improves the primitives by performing a breakdown of the execution time in \cref{fig:eval:Primitive_breakdown}.
The experiments were carried out on 32\texttimes{}32 PEs, with each PE starting with 8MB of data for ReduceScatter and each PE receiving 8MB of data for AlltoAll, AllReduce, and AllGather.
Due to highly overlapping operations, we use an approach similar to CPIstack~\cite{cpistacklike} for splitting the real system measurements.
For all primitives, the host memory access overhead is completely removed using \im.
This benefits the speedup by a geomean of 2.9\texttimes{}, overhauling the minor 4.5\% overhead of \pr.
For AlltoAll and AllGather, the overhead of domain transfer is eliminated by \cm.
Finally, the execution time of data operations such as data rearrangement and reduction from \pr allows for a more efficient operation. 

\subsection{Sensitivity Study}
\label{sec:eval:perf}

\begin{figure}
    \centering
    \setlength{\fboxsep}{0.5pt}
    \setlength{\fboxrule}{1pt}
    \includegraphics[width=0.99\columnwidth]{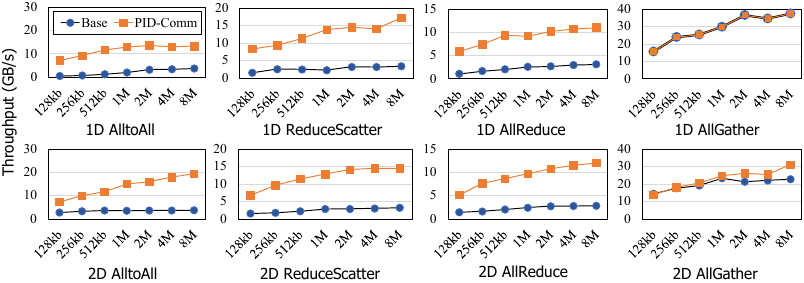}
    \caption{Performance of \thiswork primitives.} \vspace{-4mm}
    \label{fig:eval:datasize}
\end{figure}

In this section, we analyze \thiswork's performance 
on three different aspects: data size, number of PEs, and hypercube configurations.
We first compare \thiswork against the baseline for different data sizes ranging from 128K to 8M in \cref{fig:eval:datasize}.
For both hypercube configurations (1024 PEs for 1D and 32\texttimes{}32 PEs for 2D), all primitives show a similar trend of \thiswork performing better as the data size increased, reaching 2.89\texttimes{} speedup with size 8M in geomean.
The 1D AllGather is already performing effectively on the baseline, but with 2D settings, \thiswork shows improvements on large data sizes.
This is because the baseline relies on the fast broadcast function, which cannot be utilized for 2D settings. 

\begin{figure}
    \centering
    \setlength{\fboxsep}{0.5pt}
    \setlength{\fboxrule}{1pt}
    \includegraphics[width=0.97\columnwidth]{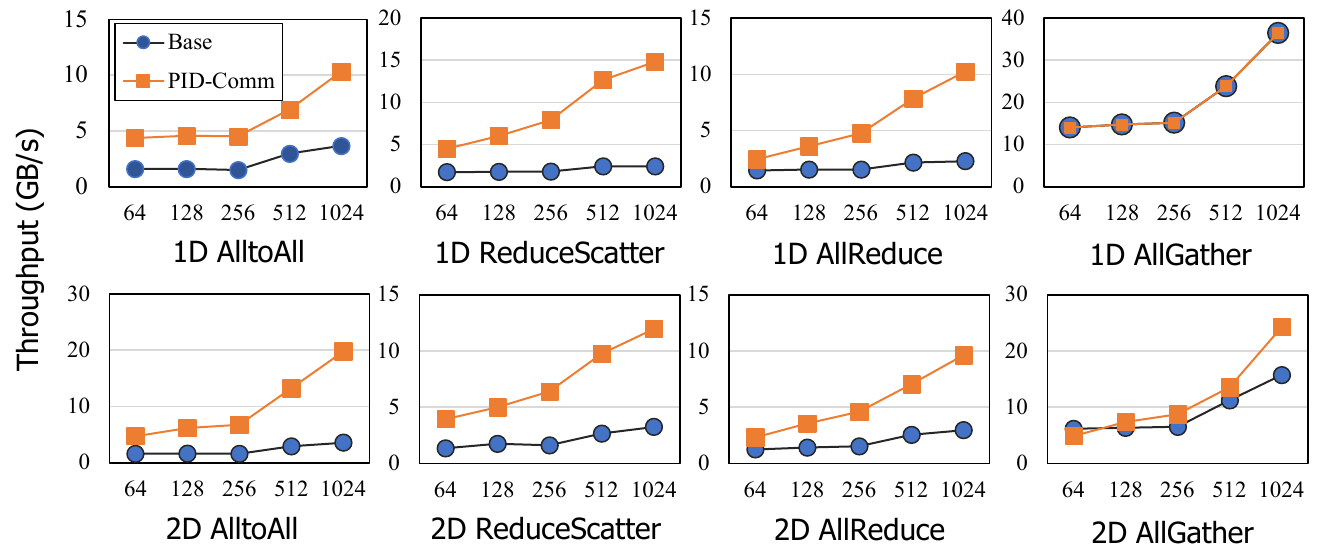}
    \caption{Performance of \thiswork primitives with varying number of PEs.}\vspace{-3mm}
    \label{fig:eval:numdpus}
\end{figure}

Next, we compare \thiswork against the baseline by changing the number of communicating PEs in \cref{fig:eval:numdpus}.
In general, \thiswork scales better as the number of PEs increase, achieving 2.36\texttimes{} to 4.20\texttimes{} more throughput as the number of PEs increased from 64 to 1024. 
For up to 256 PEs, we use one channel of UPMEM DIMMs, and increase the number of channels for settings with more PEs.
Therefore, \thiswork experiences a large throughput boost from 256 PEs to 1024 PEs. 
However, the performance of the baseline does not scale with the number of channels. 
This is because the baseline is mainly bottlenecked by the host computation rather than off-chip bandwidth, except for AllGather.
This shows that \thiswork can benefit from increasing the number of off-chip channels, a valuable resource to computing systems.

\begin{figure}
    \centering
    \setlength{\fboxsep}{0.5pt}
    \setlength{\fboxrule}{1pt}
    \includegraphics[width=0.93\columnwidth]{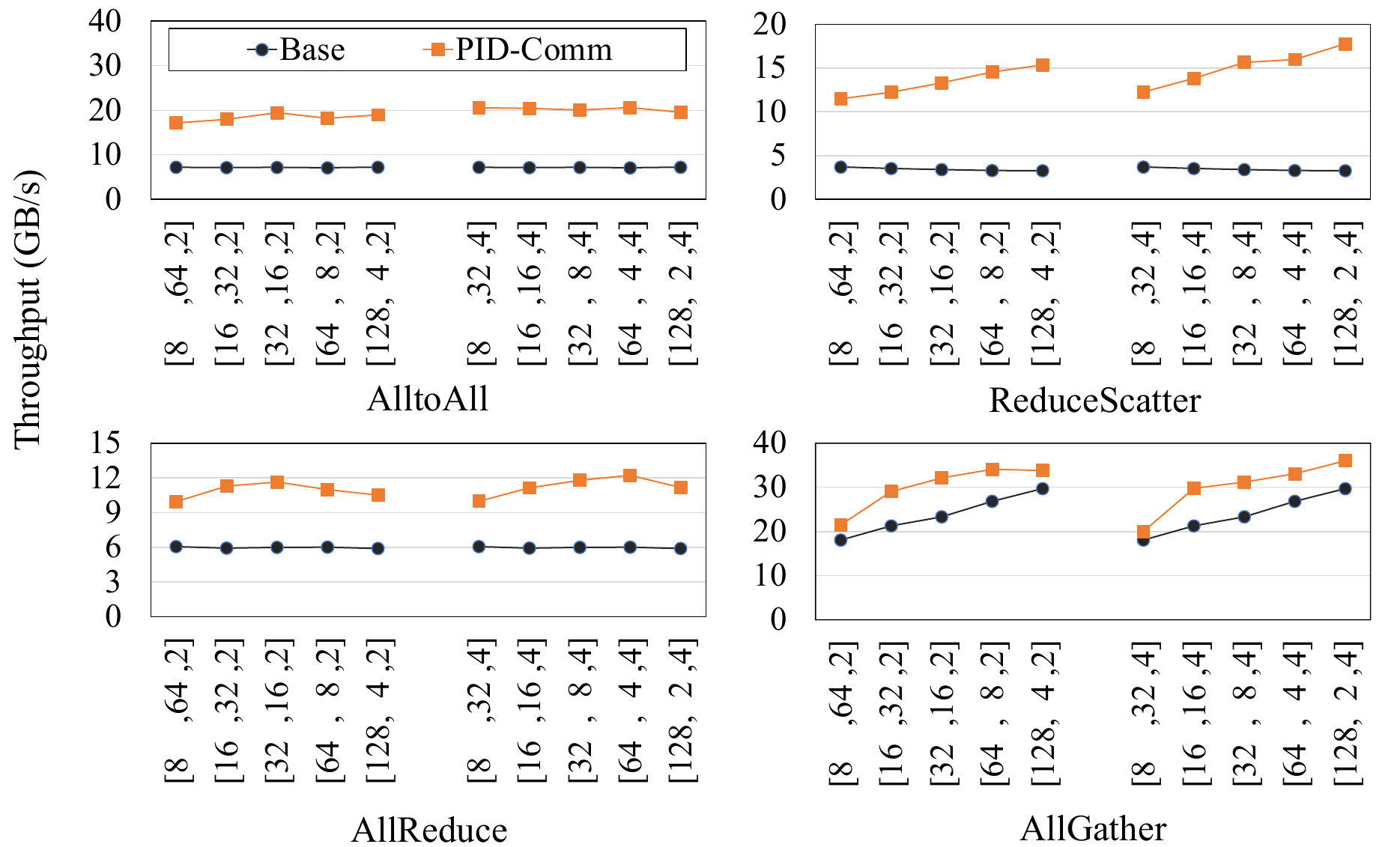}
    \caption{Speed of \thiswork with various hypercube shapes.}\vspace{-4mm}
    \label{fig:eval:hypercube}
\end{figure}

In \cref{fig:eval:hypercube}, we check the performance of \thiswork on different 3D hypercube shape configurations.
AlltoAll and AllReduce showed similar throughput for most shapes (up to 20.6 GB/s and 12.2 GB/s, respectively), while ReduceScatter and AllGather scaled up as the length of the x-axis increased (up to 17.8 GB/s and 36.1 GB/s, respectively).
The former two are not affected by the x-axis because of the fixed amount of communicating data.
Contrarily, the latter two perform better with a longer x-axis due to decreased communication. 

\begin{figure*}
    \centering
    \setlength{\fboxsep}{0.5pt}
    \setlength{\fboxrule}{1pt}
    \includegraphics[width=0.88\textwidth]{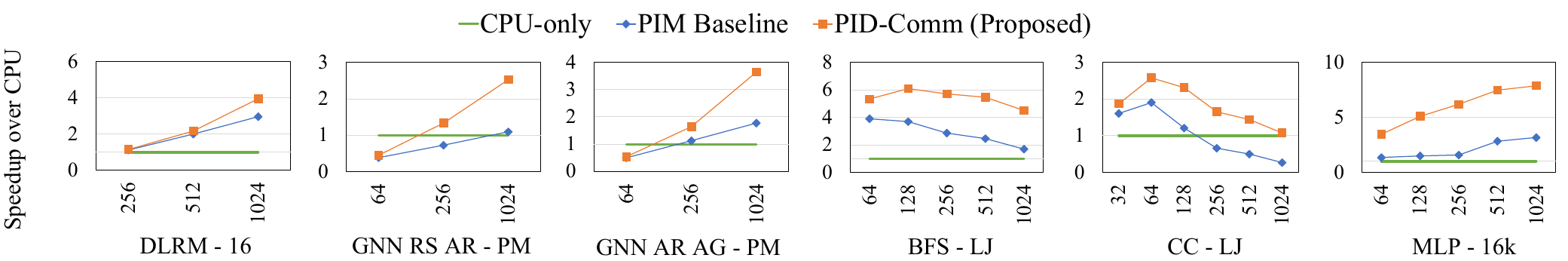}
    \caption{Performance comparison to CPU-only system with varying number of PEs.}\vspace{-4mm}
    \label{fig:eval:cpu_comp}
\end{figure*}

\subsection{Sensitivity Study on Different Word Bits}

\begin{figure}
    \centering
    \includegraphics[width=0.85\columnwidth,trim={5 5 0 0},clip]{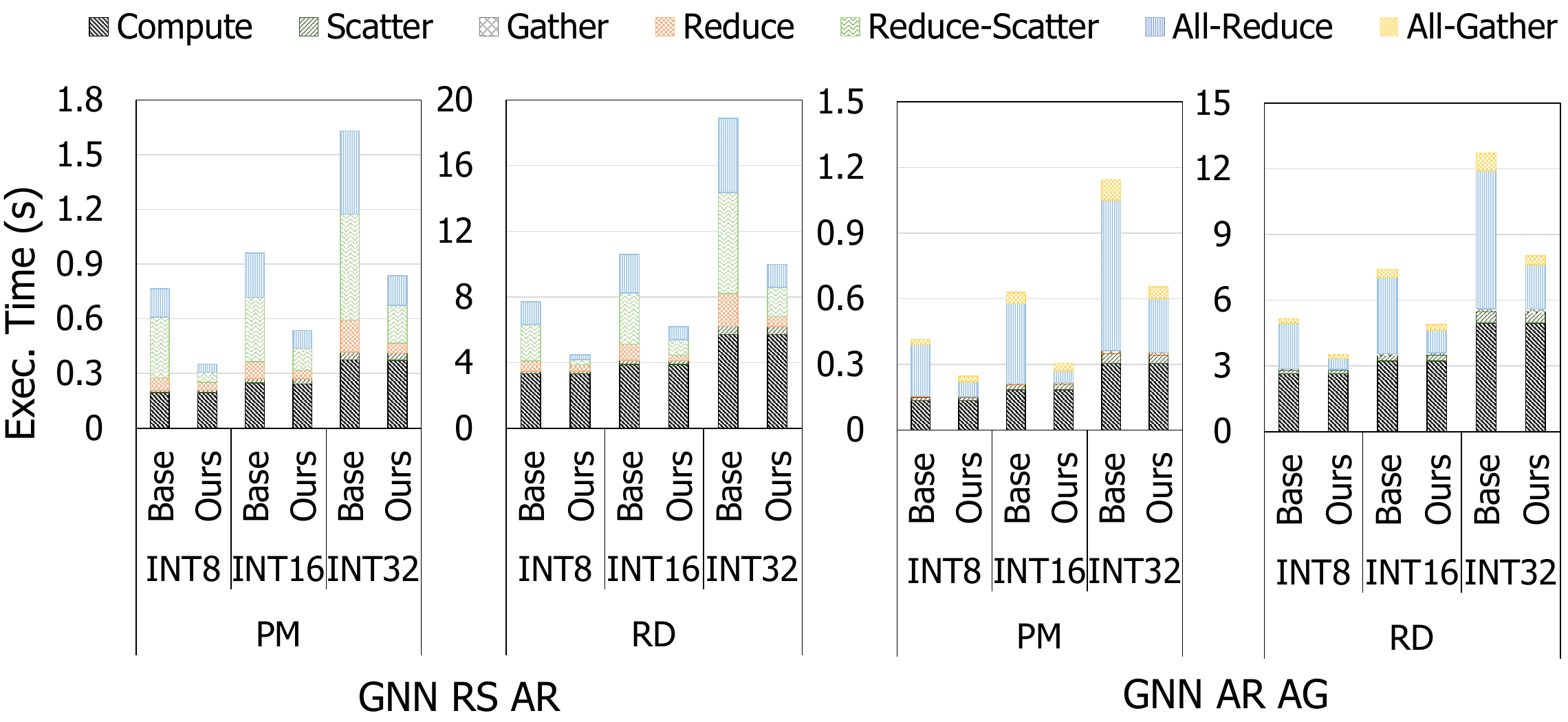}
    \caption{Sensitivity study on different word bits using GNN.}\vspace{-4mm}
    \label{fig:eval:diffbit}
\end{figure}

It is worth running a sensitivity study on different-sized data elements. 
We use GNN as a representative example and compare the performance breakdown with the baseline as in \cref{fig:eval:diffbit}.
Having a larger bit-width affects the execution time in various ways.
First, the computation takes slightly longer, mainly from more pressure on the caches (host) and scratchpad (PEs). 
In addition, communication takes longer, from increased data transfer as well as the associated data modulation.

One interesting aspect is that communication speedup is larger for 8b elements.
As mentioned in \cref{sec:general}, when the elements are 8b words, \cm can be applied to ReduceScatter and AllReduce despite the arithmetic operations within them. 
Consequently, GNNs using 8b integers achieved 1.64\texttimes{} geomean speedup compared to the baseline.
Overall, it shows that \thiswork supports multiple data granularity and provides consistent speedup over the baseline.

\begin{figure}
    \centering
    \setlength{\fboxsep}{0.5pt}
    \setlength{\fboxrule}{1pt}
    \includegraphics[width=0.85\columnwidth]{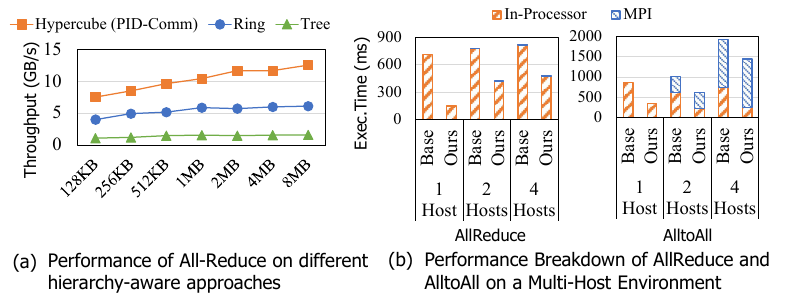}
    \caption{Performance evaluation of different hierarchies.}\vspace{-5mm}
\label{fig:eval:multihost_hier}
\end{figure}

\subsection{Comparison to CPU-only Systems}
\label{sec:eval:cpu}

To fully understand the impact of \thiswork, we compare \thiswork and the PIM baseline against the CPU-only system on the benchmark applications while varying the number of PEs from 64 to 1024 in \cref{fig:eval:cpu_comp}.
For DLRM and GNN benchmarks, some configurations are excluded as DLRM benchmarks run out of memory for small number of PEs, and GNNs require symmetric partitioning of the input graph for effective communication.
For CC, we added a 32-PE configuration to confirm the sweet spot at 64 PEs.
Overall, PIM baseline achieves 2.27\texttimes{} geomean speedup over the CPU-only system, and \thiswork achieves 4.07\texttimes{} geomean speedup.

For benchmarks with a relatively large portion of PE computation (DLRM, GNNs and MLP), the performance improved as the number of PEs increased for both \thiswork and PIM baseline. 
This is because these applications on PIM-enabled DIMMs benefit from the computational power of many PEs, effectively hiding the communication overhead. 
This led to a maximum speedup of 3.18\texttimes{} at MLP for the PIM baseline.
\thiswork additionally reduced the communication overhead and achieved a larger maximum speedup of 7.89\texttimes{} at MLP.

For other benchmarks dominated by communication between PEs (BFS, CC), some trade-off exists between the benefit of internal memory bandwidth and the overhead of inter-PE communication.
This is especially prominent in CC, where the sweet spot is around 64 PEs for both the PIM baseline and \thiswork, with \thiswork achieving a higher speedup of 2.58$\times$ over the CPU-only system.
This indicates that the configuration on PIM-based systems has to be carefully chosen, as observed in prior art~\cite{prim, pimulator}.

\vspace{2mm}
\subsection{Comparison to Other Hierarchy-Aware Approaches}

On \cref{fig:eval:multihost_hier}(a), to demonstrate the effectiveness of the virtual hypercube topology,  we compared it with popular algorithmic topologies of ring~\cite{nccl} and tree~\cite{twotree} on a 32$\times$32 two-dimensional AllReduce with all optimizations of \thiswork applied (i.e., PR, IM, and CM). 
For the former, each PE performes reduction with its physically close neighbors within the same entangled group, and then with PEs in other entangled groups.
Similarly, the latter creates reduction trees following the order of entangled group, rank, channel.

While the two compared topologies can benefit from \thiswork's optimization techniques, it lacks flexibility and wastes the available host-PIM bandwidth. 
This causes severe performance degradation, yielding at max 7.89\texttimes{} and 2.05\texttimes{} slowdown for tree and ring topology, respectively, demonstrating the need for the proposed hypercube's flexible abstraction.

\vspace{2mm}

\section{Discussion}

\subsection{\thiswork on Other PIM Architectures/Systems}
\label{sec:discussion:otherpimarch}

\begin{figure}
    \centering
    \setlength{\fboxsep}{0.5pt}
    \setlength{\fboxrule}{1pt}
    \includegraphics[width=0.82\columnwidth]{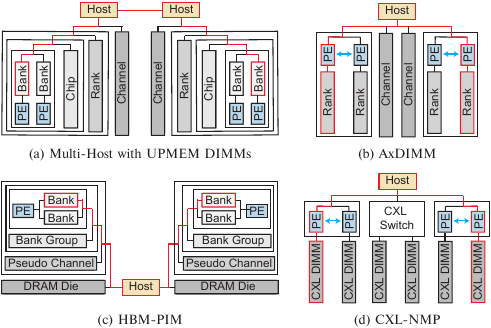}
    \caption{Communication within different PIM architectures.}\vspace{-4mm}
    \label{fig:disc:diffhard}
\end{figure}

Although \thiswork's design is rooted in 
UPMEM DIMMs, we suggest that the core concepts can be extended to other PIM hardwares unless there exists a communication medium shared by all PEs. 
Such systems can be divided into two by whether partial intermediate communication mediums (e.g. interconnect, shared buffer) exist or not. 
The former includes multi-host systems with UPMEM DIMMs and HBM-PIM~\cite{hbmpim} (\cref{fig:disc:diffhard}(a), (c)). 
The latter includes AxDIMM~\cite{axdimm} and CXL-NMP~\cite{huangfu2022beacon, sim2022computational} (\cref{fig:disc:diffhard}(b), (d)), communication mediums shown with blue arrows. 

Both can benefit from \thiswork with adjustments. 
In a multi-host setting, each host can run a collective, then run a global collective in a hierarchical manner similar to typical distributed systems~\cite{4min}.
HBM-PIM is similar to UPMEM DIMMs, where the differences include PEs being attached per two banks, and there is only a single chip.
\thiswork can be applied without \cm.
AxDIMM and CXL-NMP involve partial communication mediums. 
They require a hierarchical approach like the multi-host setting, with the connected PEs performing the first pass. 
Afterward, we can regard a group of connected PEs as a single super-PE, and the communication between is handled similarly with \thiswork.

To demonstrate this, we implemented a multi-host version of \thiswork for AllReduce and AlltoAll as in \cref{fig:eval:multihost_hier}(b). 
In the testbed, we used up to four processes, each connected to a four-rank UPMEM channel with 256 PEs with 2MB of data each, which represent an individual host.
The global communication is performed with MPI~\cite{openmpi}, whose bandwidth is controlled at 10 Gbps according to that of high-speed ethernet.
For AllReduce, the amount of MPI communication is small because the data are sent after being reduced among 256 PEs, which corresponds to 1/256 data size.
Similar trends persist in ReduceScatter whose data are sent after reduction and AllGather whose data are sent before duplication.
For AlltoAll, the overhead is relatively larger because there is no reduction or duplication. 
With more hosts, the overhead increases because more data have to travel, which roughly aligns with the known transfer cost being proportional to $N-1/N$ for $N$ hosts~\cite{mpitharkur}.
Nonetheless, \thiswork still maintains a huge advantage in throughput over the baseline, demonstrating its extendability. 

\vspace{1mm}
\subsection{Hardware Implications}
\label{sec:discussion:hw}

Due to PIM-enabled DIMM's innate hardware structure, inter-PE communication requires host involvement in linking PEs.
This burdens the host and may slow down the overall process. 
We believe this could be relieved in two directions. 

\textbf{Offloading host involvement.}
The host-side computational overhead may be relieved if we could offload the host burden to a dedicated accelerator.
A good example of such an accelerator would be the Intel Data Streaming Accelerator (DSA)~\cite{DSA}.
Unfortunately, the current version of DSA does not support essential features for \thiswork's data modulation. 
However, if future versions of DSA include more features (e.g., shifting, addition, domain transfers), we believe that DSA could fully replace the host with an even higher speedup. 

\textbf{Enabling/adding intermediate paths.}
Despite the lack of direct inter-PE paths, there already exist some physically shared mediums inside DRAM, such as internal buses between banks or off-chip channels between ranks. 
Even though those cannot be used for communication due to analog-level issues, there are suggestions to use them~\cite{rowclone, aim} or to add new connections~\cite{dimmlink}.
Results of \thiswork suggest that similar ideas are needed for more scalable PIM-enabled systems. 
Even with such hardware enhancements, the ideas of \thiswork will still be needed, as discussed in \cref{sec:discussion:otherpimarch}.

\section{Related Work}
\subsection{Processing-in-memory}
Processing-in-memory (PIM) has a long history of being studied. 
Execube~\cite{execube} is often regarded as the first PIM, followed by many approaches up to the early 2000's
~\cite{execube, yukon, flexram, diva, iram, smartmem}, resurging after a decade mainly led by 3D stacked memory~\cite{hmc}.
Many memory-intensive applications were accelerated, such as graph processing~\cite{tesseract, graphp, graphq, graphpim} or machine learning~\cite{neurocube,tetris, dracc, xnorpop, googleworkloads}.
This was followed by many directions of work such as system support~\cite{lazypim, pei}, simulators~\cite{pimulator, pimsim}, compute using memory~\cite{ambit, rowclone, computedram, fracdram}, or computing on buffer chips~\cite{axdimm, tensordimm, chameleon, nda, recnmp}, processing with non-volatile memories~\cite{pinatubo}.
Among those, near bank processors~\cite{bufcmp, gradpim, upmem, newton, aim, hbmpim, upmem, mvid} were picked up as a near-future solution.
Vendors produced real prototypes~\cite{hbmpim, aim}, and the UPMEM~\cite{upmem} is the only publically available product on the market.
Such PIM-enabled DIMMs~\cite{bufcmp, gradpim, trim, upmem} provides a practical 
solution 
for existing systems.

Many PIM techniques suffer from inter-bank communication. 
Proposed hardware solutions enhance connectivity~\cite{dimmlink, abcdimm, bioaim}, utilize bank groups~\cite{gradpim, bglp}, modify circuitry~\cite{drisa, scope, sieve}, or alter existing commands~\cite{rowclone, computedram}.
However, these will take longer to be realized, especially due to PIM-enabled DIMMs' tight limits on timing, chip area, and power. 
Alternatively, \thiswork provides a ready-to-use software-only solution to ease programming and achieve high performance on off-the-shelf PIM-attached systems.

\subsection{Collective Communication}
Collective communications are essential to parallel programming. 
\cite{tharkur,chantheory} provide effective MPI solutions for collective communication, and \cite{mamidala2008mpi, sharp, tipparaju2005optimizing, walker1994design, ring, ring2, recdouble} provide additional improvements.
\cite{mpicharacterization, laguna2019largescale} reveal that collective communications are heavily used in HPC environments.
With the growth of machine learning and its parallelization methods, 
collective primitives started to be widely used~\cite{trainingImagenetInHour, blueconnect, flexreduce, xie2022synthesizing, megatron, dlrm, sancus, overlapisca}, with library supports~\cite{nccl, sccl, mscclang, blink}, optimizations~\cite{themis, overlapisca, pidjoin, rerouting, ccube}, and architectures~\cite{innetagg, jongse, roar, innetcoll}.
To the best of our extent \thiswork is the first to provide a flexible and fast collective communication framework on PIM-enabled systems with a wide range of APIs. 


\section{Conclusion}
We propose \thiswork, a novel communication framework for PIM-enabled DIMMs.
PIMs suffer from large communication overheads due to the lack of well-defined communication models.
To address this, \thiswork defines a hypercube-based communication model and provides a highly optimized multi-instance collective communication library. 
On a real system with PIM-enabled DIMMs, \thiswork demonstrates significant speedups on multiple benchmark applications.
We believe most of the proposed ideas could be broadly applied and adopted to various PIM hardwares.

\bibliographystyle{IEEEtranS}
\bibliography{references}

\begin{thebibliography}{100}
\providecommand{\url}[1]{#1}
\csname url@samestyle\endcsname
\providecommand{\newblock}{\relax}
\providecommand{\bibinfo}[2]{#2}
\providecommand{\BIBentrySTDinterwordspacing}{\spaceskip=0pt\relax}
\providecommand{\BIBentryALTinterwordstretchfactor}{4}
\providecommand{\BIBentryALTinterwordspacing}{\spaceskip=\fontdimen2\font plus
\BIBentryALTinterwordstretchfactor\fontdimen3\font minus \fontdimen4\font\relax}
\providecommand{\BIBforeignlanguage}[2]{{%
\expandafter\ifx\csname l@#1\endcsname\relax
\typeout{** WARNING: IEEEtranS.bst: No hyphenation pattern has been}%
\typeout{** loaded for the language `#1'. Using the pattern for}%
\typeout{** the default language instead.}%
\else
\language=\csname l@#1\endcsname
\fi
#2}}
\providecommand{\BIBdecl}{\relax}
\BIBdecl

\bibitem{gapim}
N.~Abecassis, J.~Gomez~Luna, O.~Mutlu, R.~Ginosar, A.~Moisson-Franckhauser, and L.~Yavits, ``Gapim: Discovering genetic variations on a real processing-in-memory system,'' \emph{bioRxiv}, 2023.

\bibitem{tesseract}
J.~Ahn, S.~Hong, S.~Yoo, O.~Mutlu, and K.~Choi, ``{A scalable processing-in-memory accelerator for parallel graph processing},'' in \emph{ISCA}, 2015.

\bibitem{pei}
J.~Ahn, S.~Yoo, O.~Mutlu, and K.~Choi, ``{PIM-enabled instructions: a low-overhead, locality-aware processing-in-memory architecture},'' in \emph{ISCA}, 2015.

\bibitem{mcn}
M.~{Alian}, S.~W. {Min}, H.~{Asgharimoghaddam}, A.~{Dhar}, D.~K. {Wang}, T.~{Roewer}, A.~{McPadden}, O.~{O'Halloran}, D.~{Chen}, J.~{Xiong}, D.~{Kim}, W.~{Hwu}, and N.~S. {Kim}, ``{Application-transparent near-memory processing architecture with memory channel network},'' in \emph{MICRO}, 2018, pp. 802--814.

\bibitem{chameleon}
H.~Asghari-Moghaddam, Y.~H. Son, J.~H. Ahn, and N.~S. Kim, ``{Chameleon: Versatile and practical near-DRAM acceleration architecture for large memory systems},'' in \emph{MICRO}, 2016.

\bibitem{ring}
M.~Barnett, L.~Shuler, R.~van De~Geijn, S.~Gupta, D.~G. Payne, and J.~Watts, ``Interprocessor collective communication library (intercom),'' in \emph{SHPCC}, 1994, pp. 357--364.

\bibitem{googleworkloads}
A.~Boroumand, S.~Ghose, Y.~Kim, R.~Ausavarungnirun, E.~Shiu, R.~Thakur, D.~Kim, A.~Kuusela, A.~Knies, P.~Ranganathan, and O.~Mutlu, ``{Google workloads for consumer devices: Mitigating data movement bottlenecks},'' in \emph{ASPLOS}, 2018.

\bibitem{lazypim}
A.~Boroumand, S.~Ghose, M.~Patel, H.~Hassan, B.~Lucia, K.~Hsieh, K.~T. Malladi, H.~Zheng, and O.~Mutlu, ``{LazyPIM: An efficient cache coherence mechanism for processing-in-memory},'' \emph{IEEE CAL}, 2016.

\bibitem{sccl}
Z.~Cai, Z.~Liu, S.~Maleki, M.~Musuvathi, T.~Mytkowicz, J.~Nelson, and O.~Saarikivi, ``Synthesizing optimal collective algorithms,'' in \emph{PPoPP}, 2021.

\bibitem{chantheory}
E.~Chan, M.~Heimlich, A.~Purkayastha, and R.~Van De~Geijn, ``Collective communication: theory, practice, and experience,'' \emph{Concurrency and Computation: Practice and Experience}, vol.~19, no.~13, pp. 1749--1783, 2007.

\bibitem{simplepim}
J.~Chen, J.~G{\'o}mez-Luna, I.~E. Hajj, Y.~Guo, and O.~Mutlu, ``Simplepim: A software framework for productive and efficient processing-in-memory,'' \emph{arXiv preprint arXiv:2310.01893}, 2023.

\bibitem{fft}
Y.~Chen, X.~Cui, and H.~Mei, ``Large-scale fft on gpu clusters,'' in \emph{ICS}, 2010.

\bibitem{gowalla}
E.~Cho, S.~A. Myers, and J.~Leskovec, ``{Friendship and Mobility: User Movement in Location-Based Social Networks},'' in \emph{KDD}, 2011.

\bibitem{blueconnect}
M.~Cho, U.~Finkler, and D.~Kung, ``Blueconnect: Novel hierarchical all-reduce on multi-tired network for deep learning,'' in \emph{SysML}, 2016.

\bibitem{mpicharacterization}
S.~Chunduri, S.~Parker, P.~Balaji, K.~Harms, and K.~Kumaran, ``Characterization of mpi usage on a production supercomputer,'' in \emph{SC}, 2018.

\bibitem{bioaim}
J.~Cong, Z.~Fang, M.~Gill, F.~Javadi, and G.~Reinman, ``{AIM}: Accelerating computational genomics through scalable and noninvasive accelerator-interposed memory,'' in \emph{MEMSYS}, 2017.

\bibitem{avx512}
\BIBentryALTinterwordspacing
I.~Corporation, ``{Intel® AVX-512} instructions.'' [Online]. Available: \url{https://software.intel.com/en-us/articles/intel-avx-512-instructions}
\BIBentrySTDinterwordspacing

\bibitem{nccl}
\BIBentryALTinterwordspacing
N.~Corporation, ``{NVIDIA Collective Communications Library (NCCL)},'' 2017. [Online]. Available: \url{https://developer.nvidia.com/nccl}
\BIBentrySTDinterwordspacing

\bibitem{mscclang}
M.~Cowan, S.~Maleki, M.~Musuvathi, O.~Saarikivi, and Y.~Xiong, ``Mscclang: Microsoft collective communication language,'' in \emph{ASPLOS}, 2023.

\bibitem{dracc}
Q.~Deng, L.~Jiang, Y.~Zhang, M.~Zhang, and J.~Yang, ``{DrAcc}: a dram based accelerator for accurate cnn inference,'' in \emph{DAC}, 2018.

\bibitem{upmem}
F.~Devaux, ``{The true Processing In Memory accelerator},'' in \emph{HCS}, 2019.

\bibitem{diva}
J.~Draper, J.~Chame, M.~Hall, C.~Steele, T.~Barrett, J.~LaCoss, J.~Granacki, J.~Shin, C.~Chen, C.~W. Kang, I.~Kim, and G.~Daglikoca, ``{The architecture of the DIVA processing-in-memory chip},'' in \emph{ICS}, 2002.

\bibitem{cpistacklike}
P.~G. Emma, ``Understanding some simple processor-performance limits,'' \emph{IBM journal of Research and Development}, vol.~41, no.~3, pp. 215--232, 1997.

\bibitem{nda}
A.~Farmahini-Farahani, J.~H. Ahn, K.~Morrow, and N.~S. Kim, ``{NDA}: Near-dram acceleration architecture leveraging commodity dram devices and standard memory modules,'' in \emph{HPCA}, 2015.

\bibitem{computedram}
F.~Gao, G.~Tziantzioulis, and D.~Wentzlaff, ``{ComputeDRAM}: In-memory compute using off-the-shelf drams,'' in \emph{MICRO}, 2019.

\bibitem{fracdram}
F.~Gao, G.~Tziantzioulis, and D.~Wentzlaff, ``{FracDRAM: Fractional values in off-the-shelf DRAM},'' in \emph{MICRO}, 2022.

\bibitem{tetris}
M.~Gao, J.~Pu, X.~Yang, M.~Horowitz, and C.~Kozyrakis, ``{Tetris: Scalable and efficient neural network acceleration with 3d memory},'' in \emph{ASPLOS}, 2017.

\bibitem{sparsep}
C.~Giannoula, I.~Fernandez, J.~G. Luna, N.~Koziris, G.~Goumas, and O.~Mutlu, ``Sparsep: Towards efficient sparse matrix vector multiplication on real processing-in-memory architectures,'' in \emph{PMACS}, 2022.

\bibitem{prim}
J.~G{\'o}mez-Luna, I.~El~Hajj, I.~Fernandez, C.~Giannoula, G.~F. Oliveira, and O.~Mutlu, ``Benchmarking a new paradigm: Experimental analysis and characterization of a real processing-in-memory system,'' \emph{IEEE Access}, vol.~10, pp. 52\,565--52\,608, 2022.

\bibitem{upmemtraining}
J.~G{\'o}mez-Luna, Y.~Guo, S.~Brocard, J.~Legriel, R.~Cimadomo, G.~F. Oliveira, G.~Singh, and O.~Mutlu, ``An experimental evaluation of machine learning training on a real processing-in-memory system,'' \emph{arXiv preprint arXiv:2207.07886}, 2022.

\bibitem{trainingImagenetInHour}
P.~Goyal, P.~Doll{\'a}r, R.~Girshick, P.~Noordhuis, L.~Wesolowski, A.~Kyrola, A.~Tulloch, Y.~Jia, and K.~He, ``{Accurate, Large Minibatch SGD: Training Imagenet in 1 Hour},'' \emph{arXiv preprint arXiv:1706.02677}, 2017.

\bibitem{sharp}
R.~L. Graham, D.~Bureddy, P.~Lui, H.~Rosenstock, G.~Shainer, G.~Bloch, D.~Goldenerg, M.~Dubman, S.~Kotchubievsky, V.~Koushnir \emph{et~al.}, ``Scalable hierarchical aggregation protocol (sharp): A hardware architecture for efficient data reduction,'' in \emph{COMHPC}, 2016.

\bibitem{openmpi}
R.~L. Graham, T.~S. Woodall, and J.~M. Squyres, ``Open mpi: A flexible high performance mpi,'' in \emph{Parallel Processing and Applied Mathematics: 6th International Conference, PPAM 2005, Pozna{\'n}, Poland, September 11-14, 2005, Revised Selected Papers 6}.\hskip 1em plus 0.5em minus 0.4em\relax Springer, 2006, pp. 228--239.

\bibitem{reddit}
W.~Hamilton, Z.~Ying, and J.~Leskovec, ``{Inductive representation learning on large graphs},'' \emph{Advances in neural information processing systems}, 2017.

\bibitem{newton}
M.~He, C.~Song, I.~Kim, C.~Jeong, S.~Kim, I.~Park, M.~Thottethodi, and T.~Vijaykumar, ``{Newton: A DRAM-maker’s accelerator-in-memory (AiM) architecture for machine learning},'' in \emph{MICRO}, 2020.

\bibitem{huangfu2022beacon}
W.~Huangfu, K.~T. Malladi, A.~Chang, and Y.~Xie, ``Beacon: Scalable near-data-processing accelerators for genome analysis near memory pool with the cxl support,'' in \emph{2022 55th IEEE/ACM International Symposium on Microarchitecture (MICRO)}.\hskip 1em plus 0.5em minus 0.4em\relax IEEE, 2022, pp. 727--743.

\bibitem{pimulator}
B.~Hyun, T.~Kim, D.~Lee, and M.~Rhu, ``Pathfinding future pim architectures by demystifying a commercial pim technology,'' \emph{arXiv preprint arXiv:2308.00846}, 2023.

\bibitem{gloo}
\BIBentryALTinterwordspacing
F.~Inc., ``{Gloo: Collective Communications Library},'' 2020. [Online]. Available: \url{https://github.com/facebookincubator/gloo}
\BIBentrySTDinterwordspacing

\bibitem{ddr4}
JEDEC, ``{DDR4 SDRAM} specification,'' 2020, \url{https://www.jedec.org/sites/default/files/docs/JESD79-4.pdf}.

\bibitem{4min}
X.~Jia, S.~Song, W.~He, Y.~Wang, H.~Rong, F.~Zhou, L.~Xie, Z.~Guo, Y.~Yang, L.~Yu, T.~Chen, G.~Hu, S.~Shi, and X.~Chu, ``{Highly scalable deep learning training system with mixed-precision: Training imagenet in four minutes},'' \emph{NeurIPS}, 2018.

\bibitem{xnorpop}
L.~Jiang, M.~Kim, W.~Wen, and D.~Wang, ``{XNOR-POP}: A processing-in-memory architecture for binary convolutional neural networks in wide-io2 {DRAM}s,'' in \emph{ISLPED}, 2017.

\bibitem{flexram}
Y.~Kang, W.~Huang, S.-M. Yoo, D.~Keen, Z.~Ge, V.~Lam, P.~Pattnaik, and J.~Torrellas, ``{FlexRAM: toward an advanced intelligent memory system},'' in \emph{ICCD}, 1999.

\bibitem{recnmp}
L.~Ke, U.~Gupta, B.~Y. Cho, D.~Brooks, V.~Chandra, U.~Diril, A.~Firoozshahian, K.~Hazelwood, B.~Jia, H.-H.~S. Lee \emph{et~al.}, ``{Recnmp: Accelerating personalized recommendation with near-memory processing},'' in \emph{ISCA}, 2020.

\bibitem{axdimm}
L.~Ke, X.~Zhang, J.~So, J.-G. Lee, S.-H. Kang, S.~Lee, S.~Han, Y.~Cho, J.~H. Kim, Y.~Kwon, K.~Kim, J.~Jung, I.~Yun, S.~J. Park, H.~Park, J.~Song, J.~Cho, K.~Sohn, N.~S. Kim, and H.-H.~S. Lee, ``{Near-memory processing in action: Accelerating personalized recommendation with axdimm},'' \emph{IEEE Micro}, 2021.

\bibitem{mvid}
B.~Kim, J.~Chung, E.~Lee, W.~Jung, S.~Lee, J.~Choi, J.~Park, M.~Wi, S.~Lee, and J.~H. Ahn, ``{MViD}: Sparse matrix-vector multiplication in mobile dram for accelerating recurrent neural networks,'' \emph{IEEE TC}, 2020.

\bibitem{neurocube}
D.~Kim, J.~Kung, S.~Chai, S.~Yalamanchili, and S.~Mukhopadhyay, ``{Neurocube: A programmable digital neuromorphic architecture with high-density 3D memory},'' \emph{ISCA}, 2016.

\bibitem{gradpim}
H.~Kim, H.~Park, T.~Kim, K.~Cho, E.~Lee, S.~Ryu, H.-J. Lee, K.~Choi, and J.~Lee, ``{GradPIM: A practical processing-in-DRAM architecture for gradient descent},'' in \emph{HPCA}, 2021.

\bibitem{yukon}
G.~Kirsch, ``{Active Memory: Micron' s Yukon},'' in \emph{IPDPS}, 2003.

\bibitem{innetcoll}
B.~Klenk, N.~Jiang, G.~Thorson, and L.~Dennison, ``An in-network architecture for accelerating shared-memory multiprocessor collectives,'' in \emph{ISCA}, 2020.

\bibitem{execube}
P.~M. Kogge, ``{EXECUBE-A new architecture for scaleable MPPs},'' in \emph{ICPP}, 1994.

\bibitem{DSA}
R.~Kuper, I.~Jeong, Y.~Yuan, J.~Hu, R.~Wang, N.~Ranganathan, and N.~S. Kim, ``A quantitative analysis and guideline of data streaming accelerator in intel 4th gen xeon scalable processors,'' \emph{arXiv preprint arXiv:2305.02480}, 2023.

\bibitem{aim}
Y.~Kwon, K.~Vladimir, N.~Kim, W.~Shin, J.~Won, M.~Lee, H.~Joo, H.~Choi, G.~Kim, B.~An, J.~Kim, J.~Lee, I.~Kim, J.~Park, C.~Park, Y.~Song, B.~Yang, H.~Lee, S.~Kim, D.~Kwon, S.~Lee, K.~Kim, S.~Oh, J.~Park, G.~Hong, D.~Ka, K.~Hwang, J.~Park, K.~Kang, J.~Kim, J.~Jeon, M.~Lee, M.~Shin, M.~Shin, J.~Cha, C.~Jung, K.~Chang, C.~Jeong, E.~Lim, I.~Park, J.~Chun, and S.~Hynix, ``{System architecture and software stack for GDDR6-AiM},'' in \emph{HCS}, 2022.

\bibitem{tensordimm}
Y.~Kwon, Y.~Lee, and M.~Rhu, ``{TensorDIMM}: A practical near-memory processing architecture for embeddings and tensor operations in deep learning,'' in \emph{MICRO}, 2019.

\bibitem{criteo}
\BIBentryALTinterwordspacing
C.~Labs, ``Criteolabs kaggle display advertising challenge dataset.'' [Online]. Available: \url{http://labs. criteo.com/2014/02/download-kaggle-display-advertising-challenge-dataset/}
\BIBentrySTDinterwordspacing

\bibitem{laguna2019largescale}
I.~Laguna, R.~Marshall, K.~Mohror, M.~Ruefenacht, A.~Skjellum, and N.~Sultana, ``A large-scale study of mpi usage in open-source hpc applications,'' in \emph{SC}, 2019.

\bibitem{bufcmp}
J.~Lee, J.~H. Ahn, and K.~Choi, ``{Buffered compares: Excavating the hidden parallelism inside DRAM architectures with lightweight logic},'' in \emph{DATE}, 2016.

\bibitem{flexreduce}
J.~Lee, I.~Hwang, S.~Shah, and M.~Cho, ``Flexreduce: Flexible all-reduce for distributed deep learning on asymmetric network topology,'' in \emph{DAC}, 2020.

\bibitem{hbmpim}
S.~Lee, S.-h. Kang, J.~Lee, H.~Kim, E.~Lee, S.~Seo, H.~Yoon, S.~Lee, K.~Lim, H.~Shin, J.~Kim, O.~Seongil, A.~Iyer, D.~Wang, K.~Sohn, and N.~S. Kim, ``{Hardware architecture and software stack for PIM based on commercial DRAM technology: Industrial product},'' in \emph{ISCA}, 2021.

\bibitem{scope}
S.~Li, A.~O. Glova, X.~Hu, P.~Gu, D.~Niu, K.~T. Malladi, H.~Zheng, B.~Brennan, and Y.~Xie, ``Scope: A stochastic computing engine for dram-based in-situ accelerator,'' in \emph{MICRO}, 2018.

\bibitem{drisa}
S.~Li, D.~Niu, K.~T. Malladi, H.~Zheng, B.~Brennan, and Y.~Xie, ``{Drisa: A dram-based reconfigurable in-situ accelerator},'' in \emph{MICRO}, 2017.

\bibitem{pinatubo}
S.~Li, C.~Xu, Q.~Zou, J.~Zhao, Y.~Lu, and Y.~Xie, ``{Pinatubo: A processing-in-memory architecture for bulk bitwise operations in emerging non-volatile memories},'' in \emph{DAC}, 2016.

\bibitem{jongse}
Y.~Li, J.~Park, M.~Alian, Y.~Yuan, Z.~Qu, P.~Pan, R.~Wang, A.~Schwing, H.~Esmaeilzadeh, and N.~S. Kim, ``A network-centric hardware/algorithm co-design to accelerate distributed training of deep neural networks,'' in \emph{MICRO}, 2018.

\bibitem{pidjoin}
C.~Lim, S.~Lee, J.~Choi, J.~Lee, S.~Park, H.~Kim, J.~Lee, and Y.~Kim, ``Design and analysis of a processing-in-dimm join algorithm: A case study with upmem dimms,'' in \emph{SIGMOD}, 2023.

\bibitem{smartmem}
K.~Mai, T.~Paaske, N.~Jayasena, R.~Ho, W.~Dally, and M.~Horowitz, ``{Smart Memories: a modular reconfigurable architecture},'' in \emph{ISCA}, 2000.

\bibitem{mamidala2008mpi}
A.~R. Mamidala, R.~Kumar, D.~De, and D.~K. Panda, ``Mpi collectives on modern multicore clusters: Performance optimizations and communication characteristics,'' in \emph{CCGRID}, 2008.

\bibitem{dlrmisca}
D.~Mudigere, Y.~Hao, J.~Huang, Z.~Jia, A.~Tulloch, S.~Sridharan, X.~Liu, M.~Ozdal, J.~Nie, J.~Park \emph{et~al.}, ``Software-hardware co-design for fast and scalable training of deep learning recommendation models,'' in \emph{ISCA}, 2022.

\bibitem{graphpim}
L.~Nai, R.~Hadidi, J.~Sim, H.~Kim, P.~Kumar, and H.~Kim, ``{Graphpim: Enabling instruction-level pim offloading in graph computing frameworks},'' in \emph{HPCA}, 2017.

\bibitem{dlrm}
M.~Naumov, D.~Mudigere, H.-J.~M. Shi, J.~Huang, N.~Sundaraman, J.~Park, X.~Wang, U.~Gupta, C.-J. Wu, A.~G. Azzolini \emph{et~al.}, ``Deep learning recommendation model for personalization and recommendation systems,'' \emph{arXiv preprint arXiv:1906.00091}, 2019.

\bibitem{transpimlib}
G.~F. Oliveira, J.~G{\'o}mez-Luna, M.~Sadrosadati, Y.~Guo, and O.~Mutlu, ``Transpimlib: Efficient transcendental functions for processing-in-memory systems,'' in \emph{ISPASS}, 2023.

\bibitem{trim}
J.~Park, B.~Kim, S.~Yun, E.~Lee, M.~Rhu, and J.~H. Ahn, ``{Trim: Enhancing processor-memory interfaces with scalable tensor reduction in memory},'' in \emph{MICRO}, 2021.

\bibitem{ring2}
P.~Patarasuk and X.~Yuan, ``Bandwidth optimal all-reduce algorithms for clusters of workstations,'' \emph{JPDC}, vol.~69, no.~2, pp. 117--124, 2009.

\bibitem{iram}
D.~Patterson, T.~Anderson, N.~Cardwell, R.~Fromm, K.~Keeton, C.~Kozyrakis, R.~Thomas, and K.~Yelick, ``{A case for intelligent RAM},'' \emph{IEEE Micro}, 1997.

\bibitem{hmc}
J.~T. Pawlowski, ``Hybrid memory cube ({HMC}),'' in \emph{HCS}, 2011.

\bibitem{sancus}
J.~Peng, Z.~Chen, Y.~Shao, Y.~Shen, L.~Chen, and J.~Cao, ``{Sancus: Staleness-Aware Communication-Avoiding Full-Graph Decentralized Training in Large-Scale Graph Neural Networks},'' \emph{pVLDB}, 2022.

\bibitem{recdouble}
R.~Rabenseifner, ``Optimization of collective reduction operations,'' in \emph{ICCS}, 2004, pp. 1--9.

\bibitem{zero}
S.~Rajbhandari, J.~Rasley, O.~Ruwase, and Y.~He, ``{ZeRO: Memory Optimizations Toward Training Trillion Parameter Models},'' in \emph{SC}, 2020.

\bibitem{rerouting}
K.~Ranganath, A.~Abdolrashidi, S.~L. Song, and D.~Wong, ``Speeding up collective communications through inter-gpu re-routing,'' \emph{IEEE CAL}, vol.~18, no.~2, pp. 128--131, 2019.

\bibitem{overlapisca}
S.~Rashidi, M.~Denton, S.~Sridharan, S.~Srinivasan, A.~Suresh, J.~Nie, and T.~Krishna, ``Enabling compute-communication overlap in distributed deep learning training platforms,'' in \emph{ISCA}, 2021.

\bibitem{themis}
S.~Rashidi, W.~Won, S.~Srinivasan, S.~Sridharan, and T.~Krishna, ``Themis: A network bandwidth-aware collective scheduling policy for distributed training of dl models,'' in \emph{ISCA}, 2022.

\bibitem{twotree}
P.~Sanders, J.~Speck, and J.~L. Tr{\"a}ff, ``Two-tree algorithms for full bandwidth broadcast, reduction and scan,'' \emph{Parallel Computing}, vol.~35, no.~12, pp. 581--594, 2009.

\bibitem{ccube}
J.~Sanghoon, H.~Son, and J.~Kim, ``Logical/physical topology-aware collective communication in deep learning training,'' in \emph{HPCA}, 2023.

\bibitem{innetagg}
A.~Sapio, M.~Canini, C.-Y. Ho, J.~Nelson, P.~Kalnis, C.~Kim, A.~Krishnamurthy, M.~Moshref, D.~Ports, and P.~Richt{\'a}rik, ``Scaling distributed machine learning with in-network aggregation,'' in \emph{NSDI}, 2021.

\bibitem{pubmed}
P.~Sen, G.~Namata, M.~Bilgic, L.~Getoor, B.~Galligher, and T.~Eliassi-Rad, ``{Collective classification in network data},'' \emph{AI magazine}, 2008.

\bibitem{rowclone}
V.~Seshadri, Y.~Kim, C.~Fallin, D.~Lee, R.~Ausavarungnirun, G.~Pekhimenko, Y.~Luo, O.~Mutlu, P.~B. Gibbons, M.~A. Kozuch, and T.~C. Mowry, ``{RowClone: fast and energy-efficient in-DRAM bulk data copy and initialization},'' in \emph{MICRO}, 2013.

\bibitem{ambit}
V.~Seshadri, D.~Lee, T.~Mullins, H.~Hassan, A.~Boroumand, J.~Kim, M.~A. Kozuch, O.~Mutlu, P.~B. Gibbons, and T.~C. Mowry, ``{Ambit: In-memory accelerator for bulk bitwise operations using commodity DRAM technology},'' in \emph{MICRO}, 2017.

\bibitem{bglp}
W.~Shin, J.~Jang, J.~Choi, J.~Suh, and L.-S. Kim, ``{Bank-Group Level Parallelism},'' \emph{IEEE Transactions on Computers}, vol.~66, no.~8, pp. 1428--1434, 2017.

\bibitem{megatron}
M.~Shoeybi, M.~Patwary, R.~Puri, P.~LeGresley, J.~Casper, and B.~Catanzaro, ``{Megatron-LM: Training Multi-billion Parameter Language Models Using Model Parallelism},'' \emph{arXiv preprint arXiv:1909.08053}, 2019.

\bibitem{megatron-lm}
M.~Shoeybi, M.~Patwary, R.~Puri, P.~LeGresley, J.~Casper, and B.~Catanzaro, ``{Megatron-lm: Training Multi-billion Parameter Language Models Using Model Parallelism},'' \emph{arXiv preprint arXiv:1909.08053}, 2019.

\bibitem{sim2022computational}
J.~Sim, S.~Ahn, T.~Ahn, S.~Lee, M.~Rhee, J.~Kim, K.~Shin, D.~Moon, E.~Kim, and K.~Park, ``Computational cxl-memory solution for accelerating memory-intensive applications,'' \emph{IEEE Computer Architecture Letters}, vol.~22, no.~1, pp. 5--8, 2022.

\bibitem{abcdimm}
W.~Sun, Z.~Li, S.~Yin, S.~Wei, and L.~Liu, ``Abc-dimm: alleviating the bottleneck of communication in dimm-based near-memory processing with inter-dimm broadcast,'' in \emph{ISCA}, 2021.

\bibitem{mpitharkur}
R.~Thakur, R.~Rabenseifner, and W.~Gropp, ``Optimization of collective communication operations in mpich,'' \emph{The International Journal of High Performance Computing Applications}, vol.~19, no.~1, pp. 49--66, 2005.

\bibitem{tharkur}
R.~Thakur, R.~Rabenseifner, and W.~Gropp, ``Optimization of collective communication operations in {MPICH},'' \emph{The International Journal of High Performance Computing Applications}, vol.~19, no.~1, pp. 49--66, 2005.

\bibitem{tipparaju2005optimizing}
V.~Tipparaju and J.~Nieplocha, ``Optimizing all-to-all collective communication by exploiting concurrency in modern networks,'' in \emph{SC}, 2005.

\bibitem{cagnet}
A.~Tripathy, K.~Yelick, and A.~Bulu\c{c}, ``{Reducing Communication in Graph Neural Network Training},'' in \emph{SC}, 2020.

\bibitem{upmemsdk}
\BIBentryALTinterwordspacing
UPMEM, ``{UPMEM Software Development Kit (SDK)}.'' [Online]. Available: \url{https://sdk.upmem.com/}
\BIBentrySTDinterwordspacing

\bibitem{walker1994design}
D.~W. Walker, ``The design of a standard message passing interface for distributed memory concurrent computers,'' \emph{Parallel Computing}, vol.~20, no.~4, pp. 657--673, 1994.

\bibitem{blink}
G.~Wang, S.~Venkataraman, A.~Phanishayee, N.~Devanur, J.~Thelin, and I.~Stoica, ``Blink: Fast and generic collectives for distributed ml,'' in \emph{MLSys}, 2020.

\bibitem{roar}
R.~Wang, D.~Dong, F.~Lei, J.~Ma, K.~Wu, and K.~Lu, ``Roar: A router microarchitecture for in-network allreduce,'' in \emph{ICS}, 2023.

\bibitem{sieve}
L.~Wu, R.~Sharifi, M.~Lenjani, K.~Skadron, and A.~Venkat, ``Sieve: Scalable in-situ dram-based accelerator designs for massively parallel k-mer matching,'' in \emph{ISCA}, 2021.

\bibitem{xie2022synthesizing}
N.~Xie, T.~Norman, D.~Grewe, and D.~Vytiniotis, ``Synthesizing optimal parallelism placement and reduction strategies on hierarchical systems for deep learning,'' in \emph{MLSys}, 2022.

\bibitem{pimsim}
S.~Xu, X.~Chen, Y.~Wang, Y.~Han, X.~Qian, and X.~Li, ``Pimsim: A flexible and detailed processing-in-memory simulator,'' \emph{IEEE CAL}, 2018.

\bibitem{livejournal}
J.~Yang and J.~Leskovec, ``Defining and evaluating network communities based on ground-truth,'' in \emph{MDS}, 2012.

\bibitem{snf}
M.~Yoo, J.~Song, H.~Lee, J.~Lee, N.~Kim, Y.~Kim, and J.~Lee, ``Slice-and-forge: Making better use of caches for graph convolutional network accelerators,'' in \emph{PACT}, 2022.

\bibitem{graphp}
M.~Zhang, Y.~Zhuo, C.~Wang, M.~Gao, Y.~Wu, K.~Chen, C.~Kozyrakis, and X.~Qian, ``{GraphP}: Reducing communication for pim-based graph processing with efficient data partition,'' in \emph{HPCA}, 2018.

\bibitem{dimmlink}
Z.~Zhou, C.~Li, F.~Yang, and G.~Suny, ``Dimm-link: Enabling efficient inter-dimm communication for near-memory processing,'' in \emph{HPCA}, 2023.

\bibitem{graphq}
Y.~Zhuo, C.~Wang, M.~Zhang, R.~Wang, D.~Niu, Y.~Wang, and X.~Qian, ``{GraphQ}: Scalable pim-based graph processing,'' in \emph{MICRO}, 2019.

\end{thebibliography}


\end{document}